\documentclass[showpacs,prl,onecolumn,aps,superscriptaddress,preprintnumbers,nofootinbib]{revtex4}
\usepackage[T1]{fontenc}
\usepackage[latin9]{inputenc}
\usepackage{amsmath,amssymb}
\usepackage{epsfig}
\usepackage{graphicx}
\usepackage{amsmath}
\usepackage{amsfonts}
\def\slashchar#1{\setbox0=\hbox{$#1$}     		
   \dimen0=\wd0                                 	
   \setbox1=\hbox{/} \dimen1=\wd1               	
   \ifdim\dimen0>\dimen1                        	
      \rlap{\hbox to \dimen0{\hfil/\hfil}}      	
      #1                                        	
   \else                                        	
      \rlap{\hbox to \dimen1{\hfil$#1$\hfil}}   	
      /                                         	
   \fi}

\renewcommand{\vec}{\boldsymbol}
\newcommand{\beq}{\begin{equation}}
\newcommand{\eeq}{\end{equation}}
\newcommand{\bea}{\begin{eqnarray}}
\newcommand{\eea}{\end{eqnarray}}
\newcommand{\baa}{\begin{array}}
\newcommand{\eaa}{\end{array}}

\def\eq#1{{Eq.~(\ref{#1})}}

\def\fig#1{{Fig.~\ref{#1}}}
\newcommand{\aal}{\bar{\alpha}_S}
\newcommand{\bas}{\bar{\alpha}_S}
\newcommand{\as}{\alpha_S}

\newcommand{\nn}{\nonumber}
\newcommand{\ea}{e^{(\gamma-1)\xi}}

\newcommand{\h}{\frac{1}{2}}

\newcommand{\x}{\vec{x}}

\newcommand{\Lb}{\left(}
\newcommand{\Rb}{\right)}
\def\pom{{I\!\!P}}

\newcommand{\dy}{\delta Y}

\renewcommand{\vec}[1]{\boldsymbol{#1}}

\newcommand{\gga}{\tilde{\gamma}}

\newcommand{\intee}{\int_{C_{3}}  \frac{d\gamma}{2\pi i} }

\newcommand{\xx}{x^{2}_{10}}
\newcommand{\yy}{x^{2}_{20}}

\newcommand{\sat}{Q^{2}_{s}(Y_{0},b)}

\newcommand{\nexx}{N_{el}(Y_{0};x_{10},b)}
\newcommand{\neyy}{N_{el}(Y_{0};x_{20},b)}
\newcommand{\nexy}{N_{el}(Y_{0};x_{12},b)}
\newcommand{\nexxp}{N_{\rm el}^{2}(Y_{0};x_{10},b)}
\newcommand{\nexxs}{N_{\rm el}^{3}(Y_{0};x_{10},b)}
\newcommand{\nexxt}{N_{\rm el}^{4}(Y_{0};x_{10},b)}

\newcommand{\ga}{\overline{\gamma}}

\begin{document}
\title{DGLAP evolution  for DIS diffraction production of high masses.}
\author{Carlos Contreras}
\email{carlos.contreras@usm.cl}
\affiliation{Departamento de F\'isica, Universidad T\'ecnica Federico Santa Mar\'ia, and Centro Cient\'ifico-\\
Tecnol\'ogico de Valpara\'iso, Avda. Espana 1680, Casilla 110-V, Valpara\'iso, Chile}
\author{ Eugene ~ Levin}
\email{leving@post.tau.ac.il, eugeny.levin@usm.cl}
\affiliation{Departamento de F\'isica, Universidad T\'ecnica Federico Santa Mar\'ia, and Centro Cient\'ifico-\\
Tecnol\'ogico de Valpara\'iso, Avda. Espana 1680, Casilla 110-V, Valpara\'iso, Chile}
\affiliation{Department of Particle Physics, School of Physics and Astronomy,
Raymond and Beverly Sackler
 Faculty of Exact Science, Tel Aviv University, Tel Aviv, 69978, Israel}
\author{Rodrigo Meneses}
\email{rodrigo.meneses@uv.cl}
\affiliation{Escuela de Ingenier\'\i a Civil, Facultad de Ingenier\'\i a, Universidad de Valpara\'\i so, General Cruz 222, Valpara\'\i so, Chile}
\author{  Irina Potashnikova}
\email{irina.potashnikova@usm.cl}
\affiliation{Departamento de F\'isica, Universidad T\'ecnica Federico Santa Mar\'ia, and Centro Cient\'ifico-\\
Tecnol\'ogico de Valpara\'iso, Avda. Espana 1680, Casilla 110-V, Valpara\'iso, Chile}
\date{\today}

\keywords{BFKL Pomeron,  CGC/saturation approach, impact parameter dependence
 of the scattering amplitude, solution to non-linear equation, deep inelastic
 structure function}
\pacs{ 12.38.Cy, 12.38g,24.85.+p,25.30.Hm}
\begin{abstract}
 In this paper we develop the DGLAP evolution for the system of produced gluons in the process of diffractive production in DIS, directly from the evolution equation in Color Glass Condensate approach. We  are able to describe the available experimental data with small value of the QCD coupling ($\bas \approx 0.1$). We conclude that in diffractive production, we have a dilute system of emitted gluons and in the order to describe them, we need to develop the next-to-leading order approach in perturbative QCD.

  \end{abstract}

\maketitle

\vspace{-0.5cm}
\tableofcontents







\section{Introduction.}

In this paper we continue to re-visit the process of   diffractive production in the deep inelastic 
scattering in the framework of CGC/saturation approach (see Ref.\cite{KOLEB}
 for review). In this approach the diffraction production is characterized by two new scales (two saturation momenta): $Q_s\Lb Y - Y_0,Y_0\Rb$, which describes the dense system of produced gluons (see \fig{gen}-a) and
 $Q_s\Lb Y_0\Rb$, which corresponds to the dense system of gluons that is responsible for $N_{el}$ (see \fig{gen}-b).
 The equations, that govern the emission of gluons in this process, were
   proven   long ago \cite{KOLE}(see
 also Ref.\cite{HWS,HIMST,KLW}) but,  they have not been  investigated carefully for almost two decades.
 The efforts of high energy community were concentrated on  simple models
  in which
 the diffractive production of quark-antiquark pair and one additional
 gluon has been considered (see Refs.\cite{GBKW,GOLEDD,SATMOD0,KOML,
MUSCH,MASC,MAR,KLMV}).  The successful description of the old  experimental data  led to an  elusive impression,  that it is not necessary  to investigate the dense system of produced gluons in this process. In the previous paper \cite{CLMP} we developed the saturation model, in which we describe the dense system of produced gluons. However, the comparison with the experimental data shows,  that we failed to describe the data in spite of a number of the fitting parameters in the model. Based on this experience, we wish to study the emission of gluons in the DGLAP evolution with the hope, that the experimental data at small $\beta $ (see \fig{gen})  can be interpreted as  the production of a    rather dense  system of gluons,  which is not in the saturation region,  but approaching it. In other words,  we wish to search the DGLAP evolution for  dipole sizes for which $r^2 \,Q^2_s\Lb Y - Y_0, Y_0\Rb\,\,<\,\,1$. On the other hand,
we will show that  $r^2\,Q^2_s\Lb Y_0\Rb \,\sim\,1$  contribute to the elastic amplitude,  which means that we have  to take into account the saturation of gluons in the structure of $N_{el}$ in \fig{gen}-b.  Bearing this in mind we develop the DGLAP approach in the coordinate representation directly from the equations of Ref.\cite{KOLE}.

The DGLAP evolution has been discussed (see  for example, Ref.\cite{DGLAPOLD} and reference therein) but mostly, using the  Ingelman-Schlein factorization \cite{INSC} and reducing the evolution of the diffractive structure function to the DGLAP equation for the Pomeron structure function. In this paper
we take a completely different approach based on the evolution equation for the diffractive cross section of Ref.\cite{KOLE},  in which we do not introduce the so called soft Pomeron and its structure.

The paper is organized as follows. In the next section we give a brief review of the energy evolution of the processes of the diffractive production in DIS, which have been derived in CGC/saturation approach (see Refs.\cite{KOLEB,KOLE}). In section 3 we solve these equation in double log approximation (DLA) in the kinematic region where $r^2 Q^2_s\Lb Y_0,b\Rb \ll\,1$  and show that this solution can describe the experimental data. In section 4 we continue to discuss the DLA and expand it to the region  $r^2 Q^2_s\Lb Y_0,b\Rb\, \leq \,1$. We put our main attention  on fixing the initial conditions for the DLA equation. In section 5 we present the DGLAP evolution equation for the diffractive reduced cross sections. In the conclusions we summarize our results.


\section{The CGC/saturation equations for DIS diffractive productions }

\begin{figure}[ht]
   \centering
 \leavevmode
      \includegraphics[width=13cm]{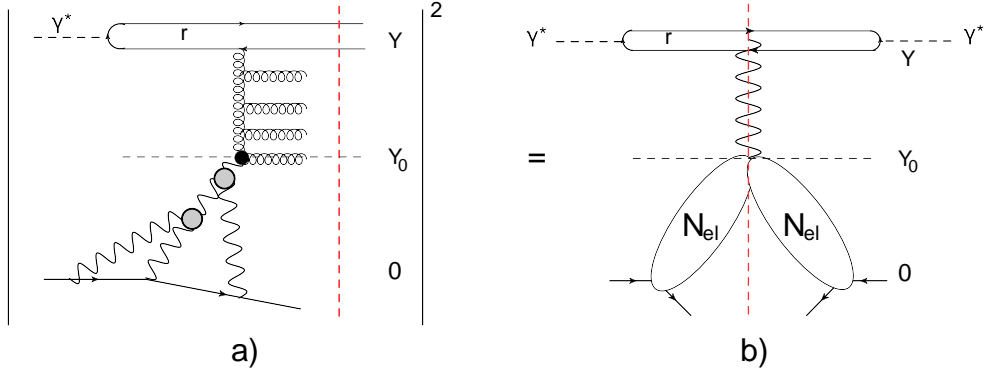}  
     \caption{The graphic representation of the processes of
 diffraction production in the region of high mass. The wavy lines denote the BFKL Pomerons.
 $Y = \ln\Lb 1/x_{Bj}\Rb, Y_0 = \ln\Lb 1/x_\pom\Rb$ and $\beta= Q^2/\Lb Q^2 + M^2_X\Rb$,  $x_\pom=\Lb Q^2 + M^2_X\Rb/s, x_{bj} = Q^2/s$ where $Q$ is the photon virtuality, $s$ the squared of energy and $ M_X$ is the produced mass.
}
\label{gen}
   \end{figure}

A sketch of the process of diffraction production in DIS is shown
 in \fig{gen}-a,  from this figure one can see that the main formula 
he form
\beq \label{EQ1}
 \sigma^{\rm diff}(Y, Y_0, Q^2)\,\,\,=\,\,\int\,\,d^2
r_{\perp} \int \,d z\,\, |\Psi^{\gamma^*}(Q^2; r_{\perp}, z)|^2
\,\,\sigma_{\rm dipole}^{diff}(r_{\perp}, Y, Y_0)\,, \eeq
where $Y = \ln\Lb 1/x_{Bj}\Rb$ and $Y_0$ is the minimum rapidity gap for
 the diffraction process (see \fig{gen}-b). In other words, we 
consider  diffraction production, in which all produced hadrons have 
rapidities larger than $Y_0$. For $\sigma_{\rm dipole}^{diff}(r_{\perp},
 Y, Y_0)$ we have a general expression
\beq \label{EQ2}
 \sigma_{\rm dipole}^{diff}(r_{\perp},Y, Y_0)
\,\,=\,\,\,\,\int\,d^2 b\,N^D(r_{\perp},Y, Y_0;\vec{b})\,, 
\eeq
where the structure of the amplitude $N^D$ is shown in  \fig{gen}-a.

For $N^D$ the evolution equation has been derived in Ref.\cite{KOLE} in the
 leading log(1/x) approximation (LLA) of perturbative
 QCD(see Ref.\cite{KOLEB} for detail and general
 description of the LLA). Hence, we hope
 to describe the experimental data only in the kinematic
 region where both $\beta$ and $x_\pom$ are very small
 ($Y - Y_0 \,=\,\ln(1/\beta) \,\gg\,1$ and $Y_0 = \ln(1/x_{\pom})
 \,\gg\,1$ are large).

The equation as has been shown in Ref.\cite{KOLE},  can be written in
 two forms. First, it turns out that for the new function
\beq \label{EQ3}
{\cal N}\Lb Y, Y_0; r_{\perp}, b \Rb\,\,\equiv\,\,2 N_{\rm el}\Lb Y;
 r_{\perp}, b \Rb \,\,-\,\,N^D(r_{\perp},x,x_0;b)
\eeq
the equation has the same form as Balitsky-Kovchegov equation \cite{BK}:
 viz.
\bea \label{EQ4}
\displaystyle{\frac{\partial {\cal N}\Lb Y, Y_0; \vec{x}_{01}, \vec{b}
 \Rb}{\partial Y}}\,&=&\,\displaystyle{\bas\int \frac{d^2  {\mathbf{x_{2}}}}{2 \pi}\frac{{\mathbf{x^2_{01}}}}{{\mathbf{x^2_{02}}}\,
{\mathbf{x^2_{12}}}}\Bigg\{ {\cal N}\Lb Y,Y_0; \vec{x}_{02},\vec{b} - \h \vec{x}_{12}\Rb +  {\cal N}\Lb Y,Y_0; \vec{x}_{12},\vec{b} - \h \vec{x}_{02}\Rb\,-\,{\cal N}\Lb Y,Y_0; \vec{x}_{01},\vec{b} \Rb}\,\nn\\
&-&\,\displaystyle{{\cal N}\Lb Y,Y_0; \vec{x}_{02},\vec{b} - \h \vec{x}_{12}\Rb {\cal N}\Lb Y,Y_0; \vec{x}_{12},\vec{b} - \h \vec{x}_{02}\Rb\Bigg\}}
\eea
with $\bas \,=\,N_c\as/\pi$ where $\as$ is QCD  coupling and $N_c$ is the number of colours.

Note, that $\vec{r} = \vec{x}_{01}$,  and the kernel of the equation
 describes the decay of   a dipole to two dipoles: 
$\vec{x}_{01}\,\to\,\vec{x}_{02}\,+\,\vec{x}_{12}$.
 The initial condition for \eq{EQ4} has the following form:
\beq \label{EQ5}
{\cal N}\Lb Y = Y_0, Y_0; \vec{x}_{01}, \vec{b} \Rb\,\,=\,\,2\,N_{\rm el}\Lb Y = Y_0; \vec{x}_{01}, \vec{b} \Rb\,-\,N^2_{\rm el}\Lb Y = Y_0; \vec{x}_{01}, \vec{b} \Rb
\eeq

Re-writing 
\eq{EQ4} as  the equation for $N^D$ we obtain the second form of
 the set of the equations:

\bea \label{EQ6}
&& \frac{   \partial N^D\Lb Y,Y_0; \vec{x}_{01},\vec{b}\Rb}{ \partial Y} \,=\\
&&\,\,\bas
 \int \,\frac{ d^2 \vec{x}_2}{2 \pi}
\frac{{\mathbf{x^2_{01}}}}{{\mathbf{x^2_{02}}}\,
{\mathbf{x^2_{12}}}}
\Bigg\{\, N^D \Lb Y,Y_0; \vec{x}_{02},\vec{b} - \h \vec{x}_{12}\Rb +  N^D\Lb Y,Y_0; \vec{x}_{12},\vec{b} - \h \vec{x}_{02}\Rb\,-\,N^D\Lb Y,Y_0; \vec{x}_{01},\vec{b} \Rb\nn\\
&& +\,\,  N^D(Y, Y_0; \vec{x}_{02},\vec{ b}- \h\vec{x}_{12} )  N^D(Y,Y_0; \vec{x}_{12},  \vec{ b}- \h\vec{x}_{02}) 
- 2 \,N^D(Y, Y_0; \vec{x}_{02},\vec{ b}- \h\vec{x}_{12} )  N_{\rm el}(Y; \vec{x}_{12},  \vec{ b}- \h\vec{x}_{02}) \nn\\
&&
\,\,- 2\,N^D(Y, Y_0; \vec{x}_{12},\vec{ b}- \h\vec{x}_{02} )  N_{\rm el}(Y; \vec{x}_{02},  \vec{ b}- \h\vec{x}_{12}\, +2\, N_{\rm el}(Y; \vec{x}_{02},\vec{ b}- \h\vec{x}_{12} )  N_{\rm el}(Y; \vec{x}_{12},  \vec{ b}- \h\vec{x}_{02})]\,\Bigg\}. \nonumber 
\eea

The initial conditions are
\beq \label{EQ7}
 N^D\Lb Y = Y_0 ,Y_0; \vec{x}_{01}, \vec{b}\Rb\,\,=\,\,N^2_{\rm el}(Y_0; \vec{x}_{01},  \vec{ b})
  \eeq
 \eq{EQ7} accounts for the production of quark-antiquark pair integrated over its mass.

A general  feature, is that the amplitude with fixed rapidity 
gap can be calculate as follows
 \beq \label{EQ8}
 n^D\Lb Y  ,\mbox{ rapidity gap}= Y_0; \vec{x}_{01},\vec{b}\Rb\,=\,- \frac{\partial  N^D\Lb Y  ,Y_0; \vec{x}_{01},\vec{b}\Rb }{\partial Y_0}\,=\,
 \frac{\partial {\cal  N}\Lb Y  ,Y_0; \vec{x}_{01},\vec{b}\Rb }{\partial Y_0}\, 
\eeq

It should be noted that the initial condition for $ n^D\Lb Y  ,\delta Y = Y - Y_0; \vec{x}_{01},\vec{b}\Rb\,=\,
\,N^2_{\rm el}(Y_0; \vec{x}_{01},  \vec{ b})\,\delta\Lb Y - Y_0\Rb$. The appearance of $\delta $-function is the artifact of the LLA, in which we only sum  contributions with large $\delta Y$. However, it has been shown \cite{GOLEDD,MAR,MUSCH} that more careful estimates of the $q \bar{q}$ production leads to smearing of the $\delta$ function and, in the first approximation,  the contribution of this production  to $n^D$ can be written as
\beq \label{EQ9}
n^D\Lb Y  ,\delta Y = Y - Y_0; \vec{x}_{01},\vec{b}\Rb\,\,=\,\,N^2_{\rm el}(Y_0; \vec{x}_{01},  \vec{ b})\,e^{ - \delta Y}
\eeq

\eq{EQ9} shows that the production of $q \bar{q}$ pairs decreases as $d \sigma/d M^2_X\,\propto\,1/M^4_x$, which corresponds to the high energy behaviour of the amplitude due to $q  \bar{q}$ exchange (see for example Ref.\cite{KOLEB}, section 3.2).

     
\begin{boldmath}
\section{ Double log approximation for the produced gluons for  $\tau_0\,=\,r^2\,Q^2_s\Lb Y_0, b\Rb\, < \,1$}
\end{boldmath}

In this section we consider the kinematic region in which
 $N_{\rm el}\Lb Y_0; \vec{x}_{01}, \vec{b}\Rb$ is in the vicinitity of
 the saturation scale $Q_s\Lb Y_0,b \Rb$, but at $x^2_{01}\,Q^2_s\Lb 
Y_0,b\Rb \, <\, 1$. As it was found in Ref. \cite{IIML,MUT} we have  geometric
 scaling behaviour in this region, and the amplitude behaves as
\beq \label{V1}
\displaystyle {N_{\rm el}\Lb Y_0; \vec{x}_{01}, \vec{b}\Rb  \,\,=\,\,{\rm c}\,\Lb x^2_{10}\,Q^2_s\Lb Y_0,b\Rb\Rb^{1 - \gamma_{cr}} }\,\,<\,\,1
\eeq
 in the leading order approximation of perturbative QCD (LOA) with $\gamma_{cr} = 0.37$.

Considering  \eq{V1}, one can see that in this kinematic region we can, 
 in general,  neglect  two terms in \eq{EQ6}:  the term which 
is
 proportional to $\Lb N^D\Rb^2$  at $Y-Y_0 \ll Y_0$, since it  is
 of the order of $N^4_{\rm el}$,  and the term which is proportional
 to $N^D N_{\rm el} \propto N^3_{\rm el}$,  while we have to keep all
 other terms.

 Therefore, the equation takes the form:

\bea \label{V3}
&& \frac{   \partial N^D\Lb Y,Y_0; \vec{x}_{01},\vec{b}\Rb}{ \partial Y} \,=\\
&&\,\bas
 \int \, \frac{d^2 \vec{x}_2}{ 2 \pi}
\frac{{\mathbf{x^2_{01}}}}{{\mathbf{x^2_{02}}}\,
{\mathbf{x^2_{12}}}}
\Bigg\{\, N^D \Lb Y,Y_0; \vec{x}_{02},\vec{b} - \h \vec{x}_{12}\Rb +  N^D\Lb Y,Y_0; \vec{x}_{12},\vec{b} - \h \vec{x}_{02}\Rb\,-\,N^D\Lb Y,Y_0; \vec{x}_{01},\vec{b} \Rb\nn\\
&&
~~~~~~~~~~~~~~~~~~~~~~~~~~~~~~+2\, N_{\rm el}(Y; \vec{x}_{02},\vec{ b}- \h\vec{x}_{12} )  N_{\rm el}(Y; \vec{x}_{12},  \vec{ b}- \h\vec{x}_{02})]\,\Bigg\}. \nonumber 
\eea

In this equation we take into account the corrections of the order
 $N^2_{el}$, but neglected  the terms of the order of $N^3_{el}$ and
 $N^4_{el}$, assuming they are small. We believe that this equation will 
allow us to take into account the correction for $N_{el} \approx 0.4 - 
0.5$.

Taking derivatives  with respect to $Y_0$, we re-write \eq{V3} for the
 amplitude $n^D\Lb Y,Y_0,\vec{x}_{01},\vec{b}\Rb$ that has been introduced
 in \eq{EQ8}. It   has the form of the linear equation:
\bea \label{V4}
&& \frac{   \partial n^D\Lb Y,Y_0; \vec{x}_{01},\vec{b}\Rb}{ \partial Y} \,=\\
&&\,\,\bas
 \int \,\frac{ d^2 \vec{x}_2}{ 2 \pi}
\frac{{\mathbf{x^2_{01}}}}{{\mathbf{x^2_{02}}}\,
{\mathbf{x^2_{12}}}}
\Bigg\{\, n^D \Lb Y,Y_0; \vec{x}_{02},\vec{b} - \h \vec{x}_{12}\Rb +  n^D\Lb Y,Y_0; \vec{x}_{12},\vec{b} - \h \vec{x}_{02}\Rb\,-\,n^D\Lb Y,Y_0; \vec{x}_{01},\vec{b} \Rb\Bigg\} \nn
\eea
The initial condition for this equation sshould be taken from \eq{EQ9} and  it has the form:

\beq \label{V5}
\displaystyle {n^D\Lb Y_0, Y_0; \vec{x}_{01}, \vec{b}\Rb  \,\,=\,\,N^2_{\rm el}\Lb Y_0; \vec{x}_{01}, \vec{b}\Rb  }  
\eeq

The elastic amplitude is:
  \beq \label{NELA}
N_{\rm el}(Y; x_{10},  b) \,\,=\,\,{\rm c}\Lb x^2_{10}\,Q^2_s\Lb Y,b\Rb\Rb^{\bar{\gamma}}\,\,\equiv\,\,{\rm c}\,\,e^{\bar{\gamma} \Lb \bas \kappa \Lb Y - Y_0\Rb \,-\, \xi\Rb}\eeq  
where $ \bar{\gamma} = 1 - \gamma_{cr}$ and  $\xi\,\equiv\,\ln\Lb 1/\Lb x^2_{10}\,Q^2_s\Lb Y_0, b\Rb\Rb\Rb$.  All other parameters of \eq{NELA} will be defined in \eq{CHI} below.

  Taking the double Mellin transform, 
  \beq \label{DM}
\displaystyle{n^D\Lb Y, Y_0, \xi, b\Rb\,\,=\,\,\, \int_{C_1}\frac{
 d \gamma d \omega}{ (2 \pi i)^2}\,\phi(\omega, \gamma)\,e^{\bas\, \omega\,
 \Lb Y- Y_0\Rb\,\,+\,\,\Lb \gamma - 1\Rb \xi}}
\eeq  
we obtain the solution to the  equation of \eq{V4} in the following form:
 \beq \label{DMS}
\displaystyle{n^D\Lb Y, Y_0, \xi, b\Rb\,\,=\,\,\, \int_{C_1}\frac{ d \gamma d \omega}{ (2 \pi i)^2}\,\frac{\phi_{in}\Lb \gamma,Y_0\Rb}{\omega - \chi\Lb \gamma\Rb}\,e^{\bas \, \omega \Lb Y  - Y_0\Rb\,\,+\,\,\Lb \gamma - 1\Rb \xi}}
\eeq  
where $\phi_{in}$ has to be determined from the initial condition of
 \eq{V5}, and it has the form
\beq \label{DM1}
\phi_{in}\Lb \gamma,Y_0\Rb\,\,=\,\,\,\frac{c^2}{\gamma - \tilde{\gamma}}
\eeq

where $\bar{\gamma}\,=\,\,1 - \gamma_{cr}$. In this paper we will use the value of $\gamma_{cr} $ which comes from the leading order estimates:    $ \gamma_{cr}\,\approx\,\,0.37$ which gives  $\bar{\gamma}\,=\,\,1 - \gamma_{cr}=0.63$ ,$ \tilde{\gamma} = -1+ 2 \gamma_{cr} =  -  0.26$.

\begin{figure}[ht]
    \centering
  \leavevmode
      \includegraphics[width=7cm]{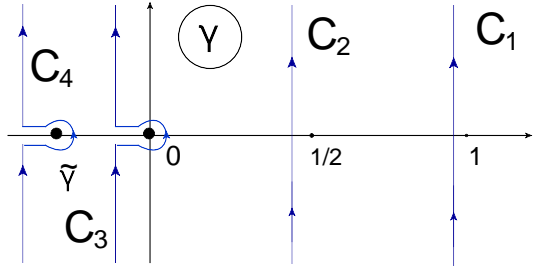}  
      \caption{The contours of integration over $\gamma$}
\label{cntr}
   \end{figure}

Therefore, the solutionhas the  form  (see \fig{gen} for  notations):

\beq \label{CR1}
\displaystyle{n^D\Lb Y, Y_0, \xi, b\Rb\,\,=\,\,{\rm c}^2
\int_{C_1}\frac{ d \gamma}{ 2 \pi i}\,
\frac{1}{ \gamma - \tilde{\gamma}}\,e^{\bas\, \chi\Lb \gamma\Rb \Lb  Y - Y_0\Rb\,\,+\,\,\Lb \gamma - 1\Rb \xi}}\eeq
with  $\chi\Lb \gamma\Rb $ and $\kappa$ are  equal to
\beq \label{CHI}
\kappa \,\,=\,\,\frac{\chi\Lb 1 - \gamma_{cr}\Rb}{1 - \gamma_{cr}}; \,\, \,\, \chi\Lb \gamma\Rb \,=\,2 \psi\Lb 1\Rb - \psi\Lb \gamma\Rb - \psi\Lb 1 - \gamma\Rb;
\eeq
 where $\psi(x) = d \ln \Gamma(x)/d x$  and $\Gamma$ is the Euler gamma function \cite{RY}.The choice of the contour of integration over $\gamma$ (see \fig{cntr})
is standard for the solution of the BFKL Pomeron, and correctly reproduces
  the calculation of the gluon emission in perturbative QCD.

The contour of integration ($C_1$) is shown in \fig{cntr}. Recalling that $x^2_{01}\,Q^2_s\Lb 
Y_0,b\Rb \, <\, 1$  and $
\xi > 0$, we can safely move this contour, and for large values of
 $\delta Y = Y - Y_0$ and $\xi$,  we can take the integral using
 the method of steepest decent. For $\as \delta Y \,\gg\,\xi$ we
 evaluate the integral by this method,   integrating along  the contour
 $C_2$ which crosses the real axis at  $\gamma$ close to $\h$.
 Here, we develop the double log(1/x) approach in which we replace the kernel of \eq{CHI} by the $\chi\Lb \gamma \Rb\,=\,1/\gamma$. 
  At $  \xi\,\gg \,\bas\,\delta Y$, we can integrate by the 
 method of steepest descend, but moving contour $C_2$  closer to $y$-axis in \fig{cntr}.
 For $\delta Y = 0$ we can close the counter over pole
 $\gamma = \bar{\gamma}$. However, for   $\delta Y \sim 1$ we
 cannot use the same method, since at $\gamma = 0$ we have singularities 
in
 the kernel $\chi(\gamma)$.

  For developing the DLA , 
 let us analyze the solution iterating the equation keeping $\bas \,\delta\, Y \,\,<\,1$. To obtain the solution as a sum of $ \Lb \delta Y\Rb^n$ contributions we need to expand
   \beq \label{V43}
 e^{ \bas \chi\Lb \gamma\Rb \Lb Y - Y_0\Rb}\,\,=\,\,\sum^\infty_{n=0} \,\frac{\Lb \bas \chi(\gamma)\,\delta Y\Rb^n}{n!} \,\,\xrightarrow{~~ \gamma \to 0~~~}\,\,\sum^\infty_{n=0} \,\frac{1}{n!} \Lb \frac{ \bas \,\delta Y}{\gamma}\Rb^n
 \eeq 
 For each term of this series,  we need to plug  in our solution,
  and integrate over $\gamma$. This integral  takes the
 following  form for the third term in \eq{CR1} for $n\,\geq\,1$:
 
 \beq \label{V44}
\oint_{C_3} \,d \gamma \,\Lb \frac{ \bas \,\delta Y}{\gamma}\Rb^n\frac{e^{ \Lb \gamma - 1\Rb \xi} }{ \gamma - \tilde{\gamma}}\,\,=\,\,
\Lb \bas \,\delta Y\Rb^n\Bigg\{\frac{1}{\tilde{\gamma}}\frac{1}{(n - 1)!} \Lb \frac{e^{ \Lb \gamma - 1\Rb \xi} }{ \gamma - \tilde{\gamma}}\Rb^{(n)}_{\gamma,\gamma \to 0}\,+\,\frac{1}{\tilde{\gamma}^n}\,e^{ \Lb\tilde{ \gamma} - 1\Rb \xi}\Bigg\}
\eeq  

For $n=0$, we only  have the contribution  of the second term in \eq{V44}.

In \eq{V43} we evaluated the integral, closing the contour over
 the singularities of the BFKL kernel, which is the pole at
 $\gamma=0$,  and over the pole $\gamma = \tilde{\gamma}$.
 The BFKL kernel also has poles  at $\gamma\,=\,- n,\, n = 1,2,3 \dots$, 
   but their contributions are exponentially suppressed with $\xi$ leading 
to
 the next twists contributions. \eq{V44} can be re-written as follows
\beq \label{V45}
\oint_{C_3} \,d \gamma \,\Lb \frac{ \bas \,\delta Y}{\gamma}\Rb^n\frac{e^{
 \Lb \gamma - 1\Rb \xi} }{ \gamma - \tilde{\gamma}}\,\,=\,\,
\Lb \frac{\bas \,\delta Y}{\tilde{\gamma}}\Rb^n\, e^{- \xi}\,\Bigg\{\frac{1}{\tilde{\gamma}}\frac{1}{(n - 1)!}\,\sum_{k=0}^{n - 1}\frac{\Lb \tilde{\gamma} \, \xi\Rb^k}{k!} \,\, +\,\,e^{\tilde{\gamma} \,\xi}\Bigg\}
\eeq
 
 The last term in \eq{V45} is the contribution at the pole $\gamma =
 \tilde{\gamma}$, while the first term is the sum of logs term giving
 the leading twist perturbative series.
 
 Bearing this in mind we can re-write \eq{CR1} in the following form
 \beq \label{CR2}
n^D\Lb Y, Y_0,\xi , b\Rb\,\,=\\
\,\,{\rm c}^2\,\Bigg\{\int_{C_1 - C_4}\frac{d \gamma}{ 2 \pi i}\,
\frac{1}{ \gamma - \tilde{\gamma}}\,e^{\bas\, \chi\Lb \gamma\Rb \Lb   Y - Y_0\Rb\,\,+\,\,\Lb \gamma - 1\Rb \xi}
\,\,+\,\,e^{ \Lb\tilde{ \gamma} - 1\Rb \xi}  \,e^{\bas \chi\Lb \tilde{\gamma}\Rb\,
\Lb Y - Y_0\Rb}\Bigg\}
\eeq
  The first term in \eq{CR2} is the difference between two integrals
 with contour $C_1$ and $C_4$, while the last term is the result of
 integrating over $\gamma$, with the contour $C_4$. The advantage
 of this form for the equation, is that it satisfies the initial
 condition, since the first term is equal to zero at $\delta Y =
 0$, and the first term generates all perturbative logs   with
 respect to the dipole sizes.
 
 In the situation when $\bas\,\delta Y\, \xi \,\geq\,1$    while $\bas \,\ll\,1$, 
 the first term reduces to the double log approximation generating
 the contribution

\beq \label{V5}
 n^D_{\rm 1-st\, term \,of\, \eq{CR2}}\Lb Y,Y_0,   r\Rb\,\,=\,\,
 {\rm  c}^2\,\frac{\bas \delta Y}{|\bar{\gamma}|}\,e^{- \xi}\,\sum^\infty_{n=1}\frac{1}{n! (n - 1)!} \Lb \bas \delta Y \xi\Rb^{n - 1}\,
 =\,{\rm c}^2 \,\frac{1}{|\bar{\gamma}|} \, \sqrt{\frac{\bas \delta Y}{\xi}}\,e^{- \xi} I_1\Lb 2 \sqrt{\bas \delta Y\,\xi}\Rb  
 \eeq
 which stems from the term with $\xi^{n - 1}$ in \eq{V45}.
 
 Finally the double log contribution takes the form:
 \begin{subequations}
 \bea 
 n^D\Lb Y,Y_0,   \xi\Rb\,\, &=&
 \,\,{\rm c}^2\Bigg\{\frac{\bas}{|\tilde{\gamma}|} \sqrt{\frac{\bas \delta Y}{\xi}}\,e^{- \xi} I_1\Lb 2 \sqrt{\bas \delta Y\,\xi}\Rb \, +\,\,e^{ \Lb\tilde{ \gamma} - 1\Rb \xi}  \,e^{- \frac{\bas}{\tilde{\gamma}}\,
\Lb Y - Y_0\Rb}\Bigg\} \label{V61}\\
&=& {\rm c}^2\frac{\bas}{|\tilde{\gamma}|} \sqrt{\frac{\bas \delta Y}{\xi}}\,e^{- \xi} I_1\Lb 2 \sqrt{\bas \delta Y\,\xi}\Rb \,\,+\,\,N^2_{el}\Lb Y; \xi\Rb\,e^{- \lambda_1\,\delta Y}\label{V62}
\eea
\end{subequations}
In the leading order estimates the value of $\lambda_1 = \bas/\tilde{\gamma} + 2\, \bas/\gamma_{cr}$. However, this term describes the quark-antiquark production whose $\delta Y$ behaviour is given by \eq{EQ9}. Based on this equation and, having in mind that the next to leading order corrections are large, we consider $\lambda_1$ as a free parameter in the description of the experimental data. We expect $\lambda_1 \approx 1 $ from \eq{EQ9}, but we will discuss this term below in more detail.

In appendix A we remove the  assumption that $\bas\,\delta Y\, \xi \gg \,1$.

     \begin{figure}[ht]
  \begin{tabular}{cc}
      \includegraphics[width=10cm]{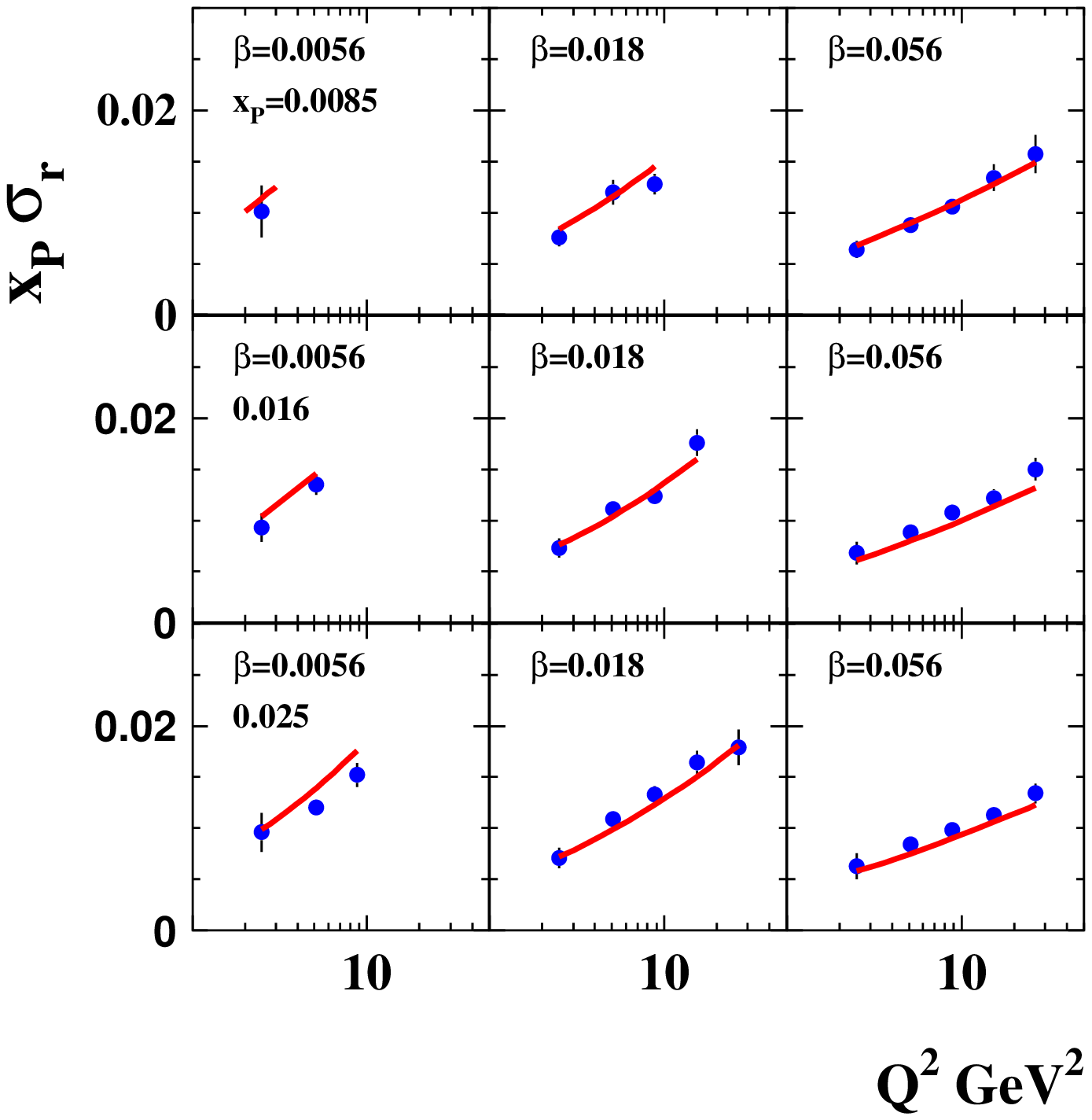} &   \includegraphics[width=10cm]{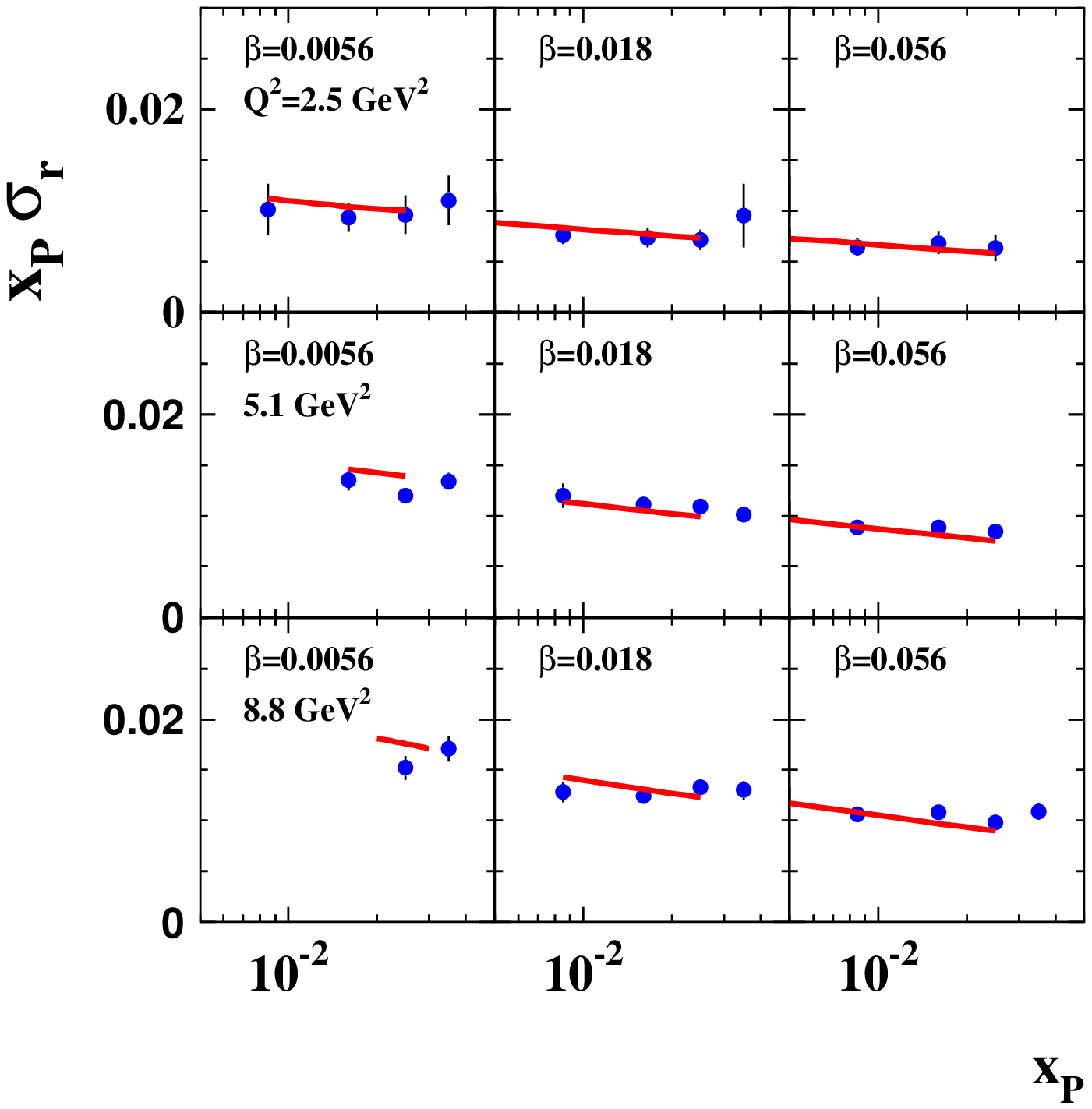}\\
      \fig{fit}-a&  \fig{fit}-b\\
      \end{tabular}
      \caption{$\sigma^{\rm diff}\Lb Y, Y_0, Q^2\Rb= x_\pom \,\sigma_r$ versus  $Q^2$ at fixed $\beta$ and $x_\pom$ (\fig{fit}-a) and  versus  $x_\pom$ at fixed $\beta$ and $Q^2$ (\fig{fit}-b). The data are taken from Ref.~\cite{HERADATA}. }
\label{fit}
   \end{figure}


Using \eq{V62} we attempted to describe  the combined set of the   inclusive diffractive cross sections measured by H1 and ZEUS collaboration at HERA~\cite{HERADATA}. The measured cross sections were expressed in terms of reduced cross sections, $\sigma_r^{D(4)}$, which is related
to the measured $ep$ cross section by
\begin{eqnarray}
\frac{{\rm d} \sigma^{ep \rightarrow eXp}}{{\rm d} \beta {\rm d} Q^2 {\rm d} \x_\pom{\rm d} t} =
\frac{4\pi\alpha^2}{\beta Q^4} \ \ \left[1-y+\frac{y^2}{2}\right] \ \
\sigma_r^{D(4)}(\beta,Q^2,x_\pom,t) \ .
\label{FIT1}
\end{eqnarray}
In the paper, the table of  $      x_\pom \sigma_r^{D(3)}(\beta,Q^2,x_\pom) = x_\pom\int d t\,\, \sigma_r^{D(4)}(\beta,Q^2,x_\pom,t) $ are presented at different values of $Q, \beta $ and $x_\pom$. This cross section is equal to  $\,\frac{Q^2}{4 \pi^2}\,\sigma^{\rm diff}\Lb Y, Y_0, Q^2\Rb$
where $\sigma^{\rm diff}\Lb Y, Y_0, Q^2\Rb$ is given by \eq{EQ1}. In the experimental data from Ref.  
  \cite{HERADATA} the integral in $t$ was performed in the region $0.09 \,\leq\,t \,\leq 0.55\, GeV^2$, while our formulae are derived for the integration in $t$ from 0 to $\infty$.  Assuming  that the $b$-dependence
of the saturation scale $Q_s\Lb Y_0,b\Rb$ is the same as was suggested in Ref. \cite{CLP} we estimate that the experimentally measured region in $t$ leads to
factor 0.52 in $ x_\pom \sigma_r^{D(3)}(\beta,Q^2,x_\pom)$.

For the fit   61 experimental points were selected which satisfy the following criteria:  $Q^2\, \leq\,26.5\,GeV^2$,  $\beta \,\leq\,0.18$ and $x_\pom \,\leq\,0.025$.  The region of $Q^2$ and $x_\pom$ was chosen from the description of the HERA data for inclusive DIS\cite{CLP}, as the specification of the region of small $x_\pom$, where the saturation model is able to fit the experimental data.
The maximal value of $\beta$ can be considered as outcome of the fit. For larger $\beta$, our approach cannot describe the data.
For this  sample we obtain the fit with $\bas = 0.119$ and $\lambda_1 = 0.6$ for  the massless light quarks and for $m_c = 1.4 \,GeV$, where $m_c$ is the mass of $c$-quark. The value of $\chi^2$ is equal 62 leading to $\chi^2/{\rm d.o.f.}\,= 1.02$.
In \fig{fit} we show   examples of the comparison of \eq{V62} with the experimental data As  expected $\lambda_1$ turns out to be close to 1.  $N_{el}$ has been taken from Ref.\cite{CLP}, where  the HERA data on $F_2$, have been described in the CGC/saturation approach. 
It is instructive to note, that the value of parameter $c$, which is needed to fit the data,  turns out to be about $0.3 - 0.4$. We believe, we can apply our approximation for  such values of $c$. Comparing with the description of the same data in our previous paper\cite{CLMP}, one can see that we obtain a much better agreement.

\begin{figure}[ht]
    \centering
  \leavevmode
      \includegraphics[width=7cm]{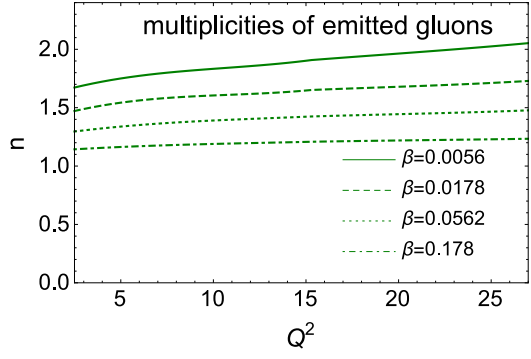}  
      \caption{ The average multiplicities of the emitted gluon at different values of $\beta$ as function of $Q^2$ using the parameters of the fit.}
\label{gmult}
   \end{figure}

 In \fig{gmult} we show the average multiplicity of the emitted gluons $n =  \sqrt{\bas \,\delta Y\,\xi}I_0\Lb 2  \sqrt{\bas \,\delta Y\,\xi} \Rb\Big{/}
 I_1\Lb 2  \sqrt{\bas \,\delta Y\,\xi} \Rb $. One can see, that we cannot restrict ourselves by calculating only one emitted gluon, as it has been discussed in Refs.\cite{GBKW,GOLEDD,SATMOD0,KOML,MUSCH,MASC,MAR,KLMV}.   On the other hand, the density is not  large to use the CGC/saturation approach for this system of gluons. As we have discussed we developed  the DLA, assuming that $\bas \,\ll\,1,\bas\,\xi \ll\,1, \,\bas\, \delta Y \,\ll \,1$ but    $ \bas \,\delta Y \,\xi \,\sim \,1$. We checked that describing the experimental data we have $ \bas\,\xi \leq 0.35$ and $ \bas \delta Y \,\leq \,0.6$ while $0.3 \,<\,\bas \delta Y \xi \,<\,2$. Based on this estimates we believe that DLA can produce a good first approximation for understanding the structure of the produced gluons. On the other hand, these estimates show that we need to go beyond the DLA.  The first corrections of this type we discuss in the appendix. Using \eq{repre} we tried to describe the data and obtain a good fit with $\chi^2 = 66$ for 61 experimental points. The values of parameters $\lambda_1=0.608$ and $\bas  =0.149$ turns out to be close to the previous fit,  showing that the corrections work in the right direction increasing the value of $\bas$. We firmly believe that the small values of the fitted $\bas$ stems from the higher order corrections, which should be taken into account.
 
 ~

 
 As has been mentioned we integrated in \eq{EQ1} only over kinematic region where $\tau_0 \,=\,r^2\,Q^2_s\Lb Y_0,b\Rb\,\leq\,1$. However, the region of $\tau_0 \,\geq 1$ does not give a negligible contribution. We checked this, integrating the second term in \eq{V62} over all $r$. We got  reasonable description of the data reaching $\chi^2 = 100$ for 61 points but  the values of parameters $\bas$ and $\lambda_1$ turn out to be quite different for  this fit : $\bas = 0.128$ and $\lambda_1= 0.95$. This shows that we need to consider the kinematic region $\tau_0 \,\leq \,1$. We  do this in the next section.
 
 ~

     
\begin{boldmath}
\section{ Double log approximation for the produced gluons for  $\tau_0\,=\,r^2\,Q^2_s\Lb Y_0, b\Rb\,\leq \,1$}
\end{boldmath}

In this section we will consider \eq{EQ6}, assuming that  the elastic amplitude $N_{el}$  is in the saturation region with $\tau_0 \,\approx\,1$. We consider that in this kinematic region $N_{el}\Lb \tau_0=1\Rb$ is not a small value, but  is of the order of 1, and, therefore, we cannot use the small sizes of $N^3_{el}$ and $N^4_{el}$ as we did deriving \eq{V4}.

 We start with specification of \eq{EQ9} for the quark-antiquark production. This process has been discussed in Refs.\cite{GBKW,GOLEDD,SATMOD0,KOML,
MUSCH,MASC,MAR,KLMV} and we briefly review the results for the completeness of the presentation.

~

~

\begin{boldmath}
\subsection{ $q \bar{q}$ production.}
\end{boldmath}

The total cross section of the quark-antiquark pair production is determined by $N^2_{\rm el}\Lb r_{\perp}, Y; \vec{b}\Rb$ and it can be found from
\eq{EQ1} with
\beq \label{QQ1}
\sigma_{\rm dipole}^{diff}(r_{\perp}, Y)\,\,=\,\,\int d^2 b\, N^2_{\rm el}\Lb r_{\perp}, Y; \vec{b}\Rb
\eeq
We need to re-write \eq{EQ1} in the following form to obtain the contribution of the $q \bar{q}$ production to the cross section with fixed produced mass $M_X$:
\beq \label{QQ2}
\sigma^{\rm diff}(Y,  Q^2)\,\,=\,\,\int d M^2_X \,\int \frac{d^2 k_T}{(2 \pi)^2}\,\delta\Lb M^2_X \,-\, \frac{k^2_T}{z (1 - z)}\Rb \Bigg( \int d^2 r_{\perp}  \Psi^{\gamma^*}\Lb Q; r_{\perp}, z\Rb N_{\rm el}\Lb r_{\perp},Y; \vec{b}\Rb \Psi^{q \bar{q}}\Lb \vec{k}_T; \vec{r}_{\perp}\Rb\Bigg)^2
\eeq
From \eq{QQ2} one can see that
\bea \label{QQ3}
&&\Lb M^2_X + Q^2\Rb \frac{d \sigma^{\rm diff}(Y, Y_M \,=\,\delta Y;  Q^2)}{d M^2_X}\,\,=\,\,
x_{\pom} \frac{d \sigma^{\rm diff}(Y, Y_M \,=\,\delta Y;  Q^2)}{d x_{\pom} }\,=\,\frac{d \sigma^{\rm diff}(Y, Y_M \,=\,\delta Y ; Q^2)}{ d Y_0}\,\,\nn\\
&&=\,\,\Lb M^2_X + Q^2\Rb\int^1_0 d z \int d^2 b \int \frac{d^2 k_T}{(2 \pi)^2}\,\delta\Lb M^2_X \,-\, \frac{k^2_T}{z (1 - z)}\Rb \Bigg( \int d^2 r_{\perp}  \Psi^{\gamma^*}\Lb Q;\vec{r}_{\perp}, z\Rb N_{\rm el}\Lb r_{\perp},Y; \vec{b}\Rb \Psi^{q \bar{q}}\Lb \vec{k}_T; \vec{r}_{\perp}\Rb\Bigg)^2 \nn\\
&&=\,\,\, \frac{Q^2}{\beta}\,\int d^2 b \int^1_0\!\!\!\!\! z (1-z) \,d z \Bigg( \int d^2 r_{\perp}  \Psi^{\gamma^*}\Lb Q;\vec{r}_{\perp}, z\Rb N_{\rm el}\Lb r_{\perp},Y; \vec{b}\Rb \Psi^{q \bar{q}}\Lb \vec{k}_T; \vec{r}_{\perp}\Rb\Bigg)^2\Bigg{|}_{k^2_T = M^2_X\,z \,(1-z)}
\eea

     \begin{figure}[ht]
    \centering
  \leavevmode
      \includegraphics[width=5cm]{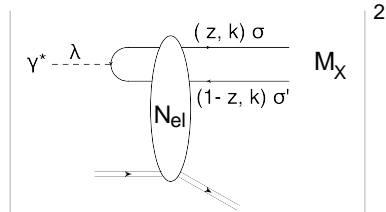}  
      \caption{The diffraction production of the quark-antiquark pair with mass $M_X$. $\lambda, \sigma$ and $\sigma'$ are photon, quark and antiquark helicities.}
\label{qq}
   \end{figure}

The wave functions of photon are  known (see Ref. \cite{KOLEB} Eqs.4.18 and 4.20) and have the following for

$\Psi^{q \bar{q}}\Lb \vec{k}_T; \vec{r}_{\perp}\Rb$ are the plane waves. Plugging \eq{PSIT} and   \eq{PSIL} into \eq{QQ3} we obtain 
 \begin{subequations}
 \bea 
\Psi_T\Lb Q; r_\perp,z\Rb\, &=&\,\frac{e Z_f}{2 \pi}\Bigg\{ (1 - \delta_{\sigma, \sigma})\,\Lb 1 - 2\,z\,-\,\sigma\,\lambda\Rb\, i a_f \frac{\vec{\epsilon}^\lambda_\perp \cdot \vec{r}_{\perp}}{ r_{\perp}} K_1\Lb r_\perp \,a_f\Rb\,+\,\delta_{\sigma\sigma'}\frac{m_f}{\sqrt{2}}\Lb 1 + \sigma  \lambda\Rb K_0\Lb r_\perp \,a_f\Rb\Bigg\};\label{PSIT}\\
N_T &=&\,\,Q^2 z (1-z)\,\,+\,\,m^2_f ~~~~\mbox{where}~~m_f ~~\mbox{ mass of the quark};\label{PSIA}\\
\Psi_L\Lb Q; r_\perp,z\Rb\, &=&\,\frac{e Z_f}{2 \pi} 2 Q z (1-z) \Lb 1 - \delta_{\sigma, \sigma'}\Rb K_0\Lb r_\perp \,a_f\Rb   \label{PSIL}
\eea
\end{subequations}
Plugging \eq{PSIT} and \eq{PSIL} into \eq{QQ3} and taking into account that $\Psi^{q\bar{q}}$ are plane waves we obtain

 \begin{subequations}
 \bea 
\,\frac{d \sigma^{\rm diff}_T(Y, Y_M \,=\,\delta Y ; Q^2)}{ d Y_0}&=&\,\frac{ 2 N_c\,\alpha_{\rm e.m.} }{ \pi^2}\sum Z^2_f \frac{Q^2}{\beta}\int d^2 b \int^1_z z(1-z) d z\Lb N_T F^2_T \,+\,m^2_f  \,F^2_L\Rb;\label{XST}\\
\,\frac{d \sigma^{\rm diff}_L(Y, Y_M \,=\,\delta Y ; Q^2)}{ d Y_0}&=&\,\frac{ 2 N_c\,\alpha_{\rm e.m.} }{ \pi^2}\sum Z^2_f\frac{4 Q^4}{\beta}\int d^2 b \int^1_z \Lb z(1-z)\Rb^3 d z \,F^2_L;\label{XSL}
\eea
\end{subequations}
where $N_T = z + (1-z)^2$ and 
 \begin{subequations}
 \bea 
F_T \,&=&\,\int \frac{r_\perp d r_\perp}{2} \,a_f\,K_1\Lb a_f\,r_{\perp}\Rb\,J_1\Lb Q \sqrt{z (1-z))} \sqrt{\frac{1 - \beta}{\beta}}\,r_\perp\Rb\,N_{\rm el}\Lb Y, r_\perp; b\Rb \label{FT};\\
F_L \,&=&\,\int \frac{r_\perp d r_\perp}{2} \,\,K_0\Lb a_f\,r_{\perp}\Rb\,J_0\Lb Q \sqrt{z (1-z)) }\sqrt{\frac{1 - \beta}{\beta}}\,r_\perp\Rb\,N_{\rm el}\Lb Y, r_\perp; b\Rb \label{FL}
\eea
\end{subequations}
 
\begin{boldmath}
\subsection{ Initial condition for evolution of $n^D$}
\end{boldmath}

 We need to return to \eq{EQ6} to find  the initial condition for the emission of the gluons in $n^D$. First, we will make the first iteration of this equation using \eq{EQ7} as the initial condition.  Plugging in the initial condition for $N^D$ from  \eq{EQ7} we obtain:
  \bea
&& \frac{   \partial }{ \partial \,\delta Y} N^D\Lb Y,Y_0; \vec{x}_{01},\vec{b}\Rb\,\,=\,\,\,\frac{\bas}{ 2 \,\pi}\,
 x^2_{10}\, \int\, \frac{d^2 \vec{x}_{02}}{\vec{x}^2_{02}\,\vec{x}^2_{12}} \Bigg\{ -\,  N^2_{\rm el} \Lb Y_0, \vec{x}_{01},\vec{b} \Rb \label{IC0}\\
 && +\,\,
\Bigg(\, N_{\rm el} \Lb Y_0, \vec{x}_{02},\vec{b} - \h \vec{x}_{12}\Rb \,+\,N_{\rm el} \Lb Y_0, \vec{x}_{12},\vec{b} - \h \vec{x}_{02}\Rb
-\, N_{\rm el} \Lb Y_0, \vec{x}_{02},\vec{b} - \h \vec{x}_{12}\Rb\, N_{\rm el} \Lb Y_0, \vec{x}_{12},\vec{b} - \h \vec{x}_{02}\Rb\Bigg)^2 \Bigg\}\nn
 \eea
 In \eq{IC0} we have two region of integrations $x_{01}\,\gg \x_{02}$ (or $x_{01}\,\gg \x_{12}$ ) and $x_{20}\,\gg\,x_{10}$,  which can lead to large logs and correspond to the singularities of the BFKL kernel. Let us first consider the region $x_{01}\,\gg \x_{02}$. In this region \eq{IC0} takes the form:
 \bea
 && \frac{   \partial }{ \partial \,\delta Y} N^D\Lb Y,Y_0; \vec{x}_{01},\vec{b}\Rb\,\,\equiv \,\,I_1\,\,=\,\,\,\h\,\bas\,
  \int^{x^2_{01}}\, \frac{d \vec{x}^2_{02}}{\vec{x}^2_{02}} \Bigg\{ -\,  N^2_{\rm el} \Lb Y_0; \vec{x}_{01},\vec{b} \Rb \label{IC00}\\
 && +\,\,
\Bigg(\, N_{\rm el} \Lb Y_0;  \vec{x}_{02},\vec{b} - \h \vec{x}_{01}\Rb \,+\,N_{\rm el} \Lb Y_0; \vec{x}_{01},\vec{b} - \h \vec{x}_{02}\Rb
-\, N_{\rm el} \Lb Y_0; \vec{x}_{02},\vec{b} - \h \vec{x}_{01}\Rb\, N_{\rm el} \Lb Y_0; \vec{x}_{01},\vec{b} - \h \vec{x}_{02}\Rb\Bigg)^2 \Bigg\}\nn
 \eea 
  From \eq{IC00} we see that the log term which is originated from the gluon reggeization (the first term in the RHS of \eq{IC0} and \eq{IC00}), cancels with the term $(\dots)^2$ and the resulting integral over $x_{02}$ leads to the contribution which is proportional to $N^2_{\rm el}\Lb Y_0, \vec{x}_{01}, \vec{b} \Rb\,\sim \,\tau_0^{2\bar{\gamma}}$ (see appendix B for more detailed  discussion of this contribution).  We show below that this contribution is much smaller  than the contribution 
 that  stem from the region of integration $x_{20}\,\gg\,x_{10}$ which generates  $\ln\Lb x^2_{10}\,Q^2_s\Lb Y_0,b\Rb\Rb$.
 The reason for this is that this term does not generate the  log contribution ($\propto \xi^n$) and can be neglected in the DLA.

  One can see that the the first iteration in this kinematic region is

  \bea
 \frac{   \partial }{ \partial \,\delta Y} \frac{N^D\Lb Y,Y_0; \vec{x}_{01},\vec{b}\Rb}{ x^2_{10}}\,&=&\,\frac{\bas}{2\,\pi}
 \int_{x_{10}} \, \frac{d^2 \vec{x}_{02}}{\vec{x}^4_{20}}
\Bigg(\,2\, N_{\rm el} \Lb Y_0; \vec{x}_{02},\vec{b} - \h \vec{x}_{02}\Rb \,-\, N^2_{\rm el} \Lb Y_0; \vec{x}_{02},\vec{b} - \h \vec{x}_{02}\Rb\Bigg)^2  \label{IC1}
 \eea
 
 The term of \eq{IC1} stems from the emission of  an extra gluon from quark - antiquark pair,  while \eq{IC00} describes the contribution of the virtual gluon to the $q \bar{q}$ production. In the general equation (see \eq{EQ6} ) this term corresponds to the reggeization of gluons.

 The first observation is that the integral in the LHS of \eq{IC1} is converged  since 
 \bea \label{IC2}
N_{\rm el} \Lb Y;  \vec{x}_{02},\vec{b} - \h \vec{x}_{02}\Rb \,\,=\,\,\left\{\begin{array}{l}\,\,\,\propto\,\,\Lb x_{02}^2 \,Q^2_s\Lb Y_0, b\Rb\Rb^{1 - \gamma_{cr}}\,\,\,\,\,\mbox{for }\,\,\, x_{02}^2 \,Q^2_s\Lb Y_0, b\Rb\,\leq \,1\,\,\mbox{with}\,\,\gamma_{cr} = 0.37;\\ \\
\,\,\,1\,\,-\,\,{\rm Const} \exp\Lb - z^2/(2 \kappa)\Rb\,\,\,\,\,\mbox{for}\,\,\, x_{02}^2 \,Q^2_s\Lb Y_0, b\Rb\,\geq\,1\,; \end{array}
\right.
\eea 
where $z \,\,=\,\,\ln\Lb x_{02}^2 \,Q^2_s\Lb Y_0, b\Rb\Rb$.

  Therefore, we can integrate in \eq{IC1} from $x_{02}=0$.   The RHS of \eq{IC1} is  $n^D\Lb Y=Y_0,Y_0,x_{01};b\Rb$ (see \eq{EQ8}). Finally, the initial condition for $n^D$ takes the form
  \bea \label{IC3}
  n^D\Lb Y=Y_0,Y_0,x_{01}; b\Rb\,\,&=&\,\,\frac{\bas}{2\,\pi} x^2_{01}\,\int_0 \, \frac{d^2 \vec{x}_{02}}{\vec{x}^4_{20}}
\Bigg(\,2\, N_{\rm el} \Lb Y_0; \vec{x}_{02},\vec{b} - \h \vec{x}_{02}\Rb \,-\, N^2_{\rm el} \Lb Y_0;  \vec{x}_{02},\vec{b} - \h \vec{x}_{02}\Rb\Bigg)^2\nn\\
&\,\,=\,\,&
\frac{\bas}{2\,\pi}\,\,x^2_{01}\,\bar{Q}^2\Lb Y_0,b\Rb
\eea
with
\bea \label{IC4}
\bar{Q}^2\Lb Y_0,b\Rb  \,\,&=&\,\,\int \frac{d^2 \vec{x}_{02}}{\vec{x}^4_{20}}
\Bigg(\,2\, N_{\rm el} \Lb Y_0; \vec{x}_{02},\vec{b} - \h \vec{x}_{02}\Rb \,-\, N^2_{\rm el} \Lb Y_0; \vec{x}_{02},\vec{b} - \h \vec{x}_{02}\Rb\Bigg)^2\,\,\nn\\
&=&\,\,
\,\,Q^2_s\Lb Y_0,b\Rb\int \frac{ \pi d \tau_0}{\tau^2_0}\Bigg(\,2\, N_{\rm el} \Lb \tau_0\Rb \,-\, N^2_{\rm el} \Lb \tau_0
\Rb\Bigg)^2 \,\,\,=\,\,\varepsilon\,\,Q^2_s\Lb Y_0,b\Rb\,
\eea
where $\varepsilon$ is a constant. Since the integral over $\tau_0$   is    convergent, the typical value of $\tau_0$ cannot  be found analytically and we used our model \cite{CLP} to estimate it. It turns out that $\varepsilon \,\approx\,\,1.2$.

It should be stressed that this contribution to $n^D$ is proportional to $x^2_{01}\,Q^2_s\Lb Y_0,b\Rb$ and it is much larger than the contribution of the order of $\tau_0^{2 \bar{\gamma}}$ that stems from the region $x^2_{01} \,\gg\,x^2_{02}$.

Finally, collecting all terms for the initial condition of $n^D$  we obtain
\beq \label{IC5}
  n^D\Lb Y=Y_0,Y_0; x_{01}, b\Rb\,\,=\,\,\frac{\bas}{2\,\pi}\,\,x^2_{01}\,\bar{Q}^2\Lb Y_0,b\Rb\, \,\,-\,\,5.62\,\,\bas \nexxp \eeq  
  where the last term we discuss in the appendix B.
  
 The first term in  \eq{IC5} was firstly derived in Ref. \cite{MAR},   considering the extra gluon emission in perturbative QCD. Here, we derived it directly from \eq{EQ6} and 
  and \eq{EQ8}.

  ~
  
  \subsection{DLA}
  
  Based on experience of the previous section we do not expect the number of emitted gluons will be large. Hence, we need to solve linear evolution equation (see \eq{V4}),  with the initial condition of \eq{IC5}. This equation in the DLA  takes the form
  \bea \label{DLA1}
 && \frac{   \partial }{ \partial \,\delta Y} \Bigg(\frac{n^D\Lb Y,Y_0; \vec{x}_{01},\vec{b}\Rb}{x^2_{01}}\Bigg)\,\,=\,\,\,\,\bas\,
  \int_{x^2_{01}}\!\!\!\frac{d \vec{x}^2_{02}}{\vec{x}^2_{02}} \,\Bigg( \frac{n^D\Lb Y,Y_0; \vec{x}_{02},\vec{b}\Rb}{x^2_{02}} \Bigg)\eea   
  
  \eq{DLA1} can be solved by the iteration with the answer:
  \beq \label{DLA2}
  n^D\Lb Y,Y_0; \vec{x}_{01},\vec{b}\Rb\,\,=\,\,\frac{\bas}{2\,\pi}\,\,x^2_{01}\,\bar{Q}^2\Lb Y_0,b\Rb \,\,I_0\Lb 2 \sqrt{\bas \,\delta Y\,\xi}\Rb\,\,=\,\,\varepsilon\,\frac{\bas}{2\,\pi}\,\,x^2_{01}\,Q_s^2\Lb Y_0,b\Rb \,\,I_0\Lb 2 \sqrt{\bas \,\delta Y\,\xi}\Rb  \eeq 
  
   The cross section has the following form
   \bea \label{DLA3}
   \sigma^{\rm diff}(Y, Y_0, Q^2)\,\,\,&=&\,\,\int\,\,d^2
r_{\perp} \int \,d z\,\, |\Psi^{\gamma^*}(Q^2; r_{\perp}, z)|^2\,\int d^2 b\,n^D_{\eq{DLA2}}\Lb Y,Y_0; \vec{x}_{01},\vec{b}\Rb\,\,\nn\\
&+&\,\,\,\Lb 1 - \,5.62\,\bas\Rb\Lb \frac{d \sigma^{\rm diff}_T(Y, Y_M \,=\,\delta Y ; Q^2)}{ d Y_0}\Bigg{|}_{\eq{XST}}\,+\,\,\,\frac{d \sigma^{\rm diff}_L(Y, Y_M \,=\,\delta Y ; Q^2)}{ d Y_0}\Bigg{|}_{\eq{XSL}}\Rb
\eea

  We compare \eq{DLA3} with the experimental data,  choosing 42 experimental points (see Ref.\cite{HERADATA}) with $\beta \,\leq\,0.056$. All other kinematic variables were the same, as in  section III, in this comparison. We obtain the description with $\chi^2 = 82$ and $\bas = 0.063$. One can see that the value of $\bas$ turns out to be smaller than in the previous section. This fact can be an  indication that the higher order corrections are essential,  that we need 
  to take into account the full DGLAP kernel for $\xi$ evolution. The high order correction we leave for the future publications,  and consider the full DGLAP approach in the next section.
  
As far as comparison with the experimental data is concerned, we consider the comparison as a good, especially since we  had only one parameter. 

   \begin{boldmath}
  \subsection{$n^D$ in the vicinity of the saturation scale $Q_s\Lb Y - Y_0, b\Rb$}
  \end{boldmath}
 \eq{DLA2} can be re-written at large $\delta Y$ as
 
 \bea \label{DLA4}
  n^D\Lb Y,Y_0; \vec{x}_{01},\vec{b}\Rb\,\,&=&\,\varepsilon\,\frac{\bas}{2\,\pi}\,\,x^2_{01}\,Q_s^2\Lb Y_0,b\Rb \,\,I_0\Lb 2 \sqrt{\bas \,\delta Y\,\xi}\Rb\nn\\
  & \xrightarrow{\bas \,\delta Y\,\xi\,\gg\,1}\,&
   \varepsilon\,\frac{\bas}{2\,\pi} \frac{1}{\sqrt{4 \pi \sqrt{\bas \delta Y\,\xi}}}\exp\Lb 2 \sqrt{\bas \delta Y \,\xi} \,-\,\xi\Rb
   \eea
  From equation $ 2 \sqrt{\bas \delta Y \,\xi} \,-\,\xi\,\,=\,0$ we find 
  \beq \label{DLAQS}
  \xi_s \,\,=\,\,\ln\Lb\frac{1}{x^2_s\,Q_s\Lb Y_0\Rb}\Rb\,\,=\,\,4\,\bas\,\delta Y
  \eeq
   which is the equation for the saturation scale: $\xi_s = \bas \kappa \,\delta Y$, in the   case of the DLA. However, we see that the solution is not a constant, but smoothly (logarithmically) depends on $\delta Y$. Therefore, the simple equation for $Q_s = Q_0 \exp\Lb \bas \kappa\,\delta Y\Rb$ has to be corrected \cite{MUT,MUPE,KOLEB}.
  It turns out \cite{MUT} that the corrected formula for energy dependence of the saturation momentum has the following form:
  \beq \label{DLA5}
   Q^2_s\Lb \delta Y\Rb\,=\,Q^2_0 \Lb\frac{1}{\bas \delta Y}\Rb^{\frac{3}{\Lb 2\,\bar{\gamma}\Rb}}\,e^{\bas\,\kappa\,\delta Y}
   \eeq
  From \eq{DLAQS} one can see that $Q_0^2 = Q^2_s\Lb Y_0\Rb$.  \eq{DLA2} in the vicinity of the saturation scale, where $ x^2_{01} Q^2_s\Lb  \delta Y\Rb \sim 1$, gives
  
  \beq \label{DLA6}
  n^D \,\,=\,\,{\rm Const}\Lb x^2_{01}\,Q^2_s\Lb \delta Y\Rb\Rb^{\bar{\gamma}}\,\,=\,\,
  {\rm Const}\Lb x^2_{01} Q^2_s\Lb Y = 0\Rb\Rb^{\bar{\gamma}}\,\frac{e^{ \bas\,\kappa\,\bar{\gamma} Y}}{\Lb\bas Y_0\,\bas\delta Y\Rb^{3/2}}
  \eeq
  \eq{DLA6} has  simple meaning, that the scale of the dense system of emitted gluons is determined by the saturation scale $ Q_s\Lb \delta Y\Rb$, which is equal for $\delta Y = 0$ to the saturation scale of the parton system which leads to $N_{el}\Lb x_{01},Y_0\Rb$. This equation has been derived recently \cite{MUMU}\footnote{At first sight, the equation in Ref.\cite{MUMU} has an extra $\ln\Lb x^2_{01} Q^2_s\Lb Y\Rb\Rb$. However, this log also enters our formula and is  absorbed in ${\rm Const}$ in our equation. For completeness of presentation we discuss this formula in more details in appendix D.} from different and more microscopic insight in the evolution of the emitted dipoles. 
  
  ~

  ~
   
   \begin{boldmath}
  \section{DGLAP evolution for the emitted gluons and quarks at  small $\beta$}
  \end{boldmath}
  \eq{DLA1}, which sums the diagrams in the leading log approximation, guides us in  writing the DLAP evolution equations. Indeed, it shows that the physics observable,  for which we can write the DLA is $g^D\Lb Y,Y_0,x_{01}\Rb \,\,\equiv\,\,n^D\Lb Y,Y_0; \vec{x}_{01},\vec{b}\Rb\Big{/}x^2_{01}$ for which the equation   can be re-written in the form
  \beq \label{DGLAP1}
  \frac{\partial g^D\Lb Y,Y_0,x_{01}\Rb}{\partial \xi}\,\,=\,\,\bas \int^Y_{Y_0} d Y'    \, g^D\Lb Y',Y_0,x_{01}\Rb\,\,=\,\,\bas \int^{\delta Y}_{0} d\, \delta Y'    \, g^D\Lb \delta Y',Y_0,x_{01}\Rb\  \eeq
  
  \eq{DGLAP1} is written for small $\beta$ but can be easily generalized for any values of $\beta$ replacing $ d Y' = -  d \beta'/\beta'$ integration by the DGLAP kernel: viz.
  \beq \label{DGLAP2}
    \frac{\partial\, \beta G^D\Lb \beta,Y_0,\xi\Rb}{\partial \xi}  \,\,=\,\,\bas \int^1_\beta d \beta' P_{gg}\Lb \beta'\Rb\,\Lb \frac{\beta}{\beta'}\, G^D\Lb \frac{\beta}{\beta'}, \xi\Rb\Rb
    \eeq
  where $  \beta \,G\Lb \beta,Y_0,\xi\Rb\,\equiv\,n^D\Lb \beta, Y_0,\xi\Rb\,e^{ - \xi} $, which is the gluon structure function for diffractively produced gluons in coordinate representation. 
  \beq \label{DGLAP3}
   \beta \,G^D\Lb \beta,Y_0,\xi = 0 \Rb  \,\,=\,\,\frac{\bas}{2 \pi}\varepsilon \eeq

   The general form of the DGLAP \cite{DGLAP} equation takes the form:
  \beq \label{DGLAP4}
 \frac{\partial }{\partial\,\xi}  \left( \begin{array}{c}
 \beta \,G^D\Lb \beta, Y_0,\xi\Rb \\
 \beta \,\Sigma^D\Lb \beta, Y_0,\xi\Rb
\end{array} \right)\,\,  =\,\,\bas\,\int^1_\beta d \beta'\,
\left( \begin{array}{cc}
P_{gg}\Lb \beta'\Rb & P_{gf}\Lb \beta'\Rb  \\
P_{fg}\Lb \beta'\Rb & P_{ff}\Lb \beta'\Rb\end{array} \right)\left( \begin{array}{c}
 \frac{\beta}{\beta'}\,G^D\Lb \frac{\beta}{\beta'}, Y_0,\xi\Rb \\
 \frac{\beta}{\beta'} \,\Sigma^D\Lb \frac{\beta}{\beta'}, Y_0,\xi\Rb
\end{array} \right)   
\eeq

In \eq{DGLAP4} $\Sigma\Lb \beta,Y_0,\xi\Rb$ is the structure function of sea quarks and antiquarks in the coordinate representation. The initial condition in our approach uhas the form
\beq \label{DGLAP5}
\Sigma^D\Lb \beta, Y_0, \xi = 0\Rb\,\,=\,\,0
\eeq
The splitting functions are well known and can be found in any text book (see Refs.\cite{KOLEB,EKL} for example). 

We  need to use the $\omega$-representation, viz.

\beq \label{OMRE}
f\Lb \omega, \xi\Rb\,=\,\int^1_0 d \beta \beta^\omega\,f\Lb \beta, \xi\Rb \,=\,\int^\infty_0\,d Y \,e^{\omega\,Y}\Lb \beta f\Lb \beta, \xi\Rb\Rb;~~~~~~~~f\Lb \beta, \xi\Rb\,=\,\int^{\epsilon + i \infty}_{\epsilon -  i \infty}\frac{d \omega}{2 \,\pi\,i} \beta^{- \omega - 1 }\,f\Lb \omega, \xi\Rb,
\eeq
to solve \eq{DGLAP4}. In this representation the equation takes the form:
  \beq \label{DGLAP6}
 \frac{\partial }{\partial\,\xi}  \left( \begin{array}{c}
 \,G^D\Lb \omega, Y_0,\xi\Rb \\
\,\Sigma^D\Lb \omega, Y_0,\xi\Rb
\end{array} \right)\,\,  =\,\,\bas\,
\left( \begin{array}{cc}
\gamma_{gg}\Lb \omega\Rb & \gamma_{gf}\Lb \omega\Rb  \\
\gamma_{fg}\Lb \omega\Rb & \gamma_{ff}\Lb \omega\Rb\end{array} \right)\left( \begin{array}{c}
\,G^D\Lb\omega, Y_0,\xi\Rb \\
\,\Sigma^D\Lb \omega, Y_0,\xi\Rb
\end{array} \right)   
\eeq  
where $$\gamma_{i,j}\,\,=\,\,\int^1_0 d z z^\omega\,P_{i,j}\Lb z \Rb$$ and for the completeness of presentation their explicit forms in the leading order are given in the appendix C.

The solution to \eq{DGLAP6}  in the region of small $\beta$,  has been discussed in details  in Ref. \cite{EKL}. The solution to the secular(characteristic)  equation that corresponds to \eq{DGLAP6} has the following form:
\beq \label{DGLAP7}
\lambda_{\pm}\,\,=\,\,\h\Bigg\{ \gamma_{ff}\Lb \omega \Rb\,+\,\gamma_{gg}\Lb \omega \Rb   \,\,\pm\,\,\sqrt{\Lb \gamma_{ff}\Lb \omega \Rb   \,-\,\gamma_{gg}\Lb \omega \Rb\Rb^2 \,\,+\,\,4\,\gamma_{fg}\Lb \omega \Rb \,\gamma_{gf}\Lb \omega \Rb }\Bigg\}
\eeq

\begin{figure}[h]
    \centering
  \leavevmode
      \includegraphics[width=8cm]{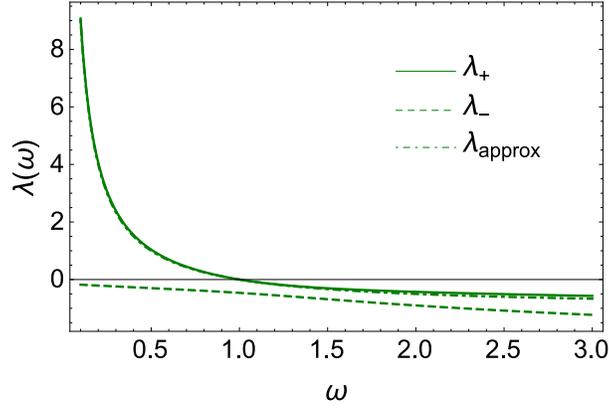}  
      \caption{$\lambda_{\pm}$ versus $\omega$ together with $\lambda_{\rm approx}\Lb \omega\Rb\,\,=\,\,\frac{1}{\omega} - 1$.}
\label{lam}
   \end{figure}

In \fig{lam} we show the dependence on $\omega$ for $\lambda_{\pm}$. One can see that only $\lambda_{+}$ is large, at the small values of $\omega$ \cite{EKL}. Indeed, the expansion of $\lambda_{\pm}$ \cite{EKL} at $\omega = 0$ gives (for $N_c = N_f =3$)

\beq \label{DGLAP8}
\lambda_{+}\Lb \omega\Rb \,\,=\,\,\frac{1}{\omega}\,\,-\,\,\frac{101}{108};~~~~~\lambda_{-}\,\,=\,\,- \frac{4}{27};
\eeq
In ref.\cite{EKL} it was noted that $\lambda_{+}\Lb \omega\Rb\,\,= \,\,\lambda_{\rm appr}\Lb \omega\Rb\,\,=\,\,1/\omega -1$ within 5\% accuracy for the values of $ \omega \leq 3$. It should be stressed that both $\lambda_{+}$ and $\lambda_{\rm approx}$ have zeros at $\omega =1$ which follow from the energy conservation.

The general form of the solution has the following form\cite{EKL}:

\beq \label{DGLAP9}
\Bigg\{\mathbf{P}_{+}\,e^{\bas\, \lambda_{+}\Lb \omega\Rb\,\xi}\,\,+\,\,\mathbf{P}_{-}\,e^{ \bas\,\lambda_{-}\Lb \omega\Rb\,\xi}\Bigg\}
\left( \begin{array}{c}
\,G\Lb\omega, Y_0,\xi = 0\Rb \\
\,\Sigma\Lb \omega, Y_0,\xi = 0\Rb
\end{array} \right)  
\eeq 
where operators $\mathbf{P}_{\pm}$ are the projectors on the eigenfunction of \eq{DGLAP6}: $\mathbf{P}_{+}\,+\,
 \mathbf{P}_{-}\,\,\equiv\,\,\mathbf{I}$ where $\mathbf{I}$ is the unit operator.
 In Ref.\cite{EKL} the operators $\mathbf{P}_{\pm}$ are found in the region of small $\omega$ (small
 $ \beta$) and they have the form:
 \beq \label{OPP}
 \mathbf{P}_{+}\,\,=\,\, \left( \begin{array}{cc}
1 & \frac{C_F}{C_A}  \\
0 & 0 \end{array} \right) \,\,+\,\,\omega \, \left( \begin{array}{cc}
\rho  & -\,C_F  \\
C_F & C_A \end{array} \right) \,\,=\,\, \left( \begin{array}{cc}
1 & 4/9 \\
0 & 0 \end{array} \right) \,\,+\,\,\omega \, \left( \begin{array}{cc}
85/54  & -\,4/3  \\
4/3 & 3 \end{array} \right)\eeq
where $C_A = N_c$ ,  $\rho \,=\frac{C_F\,C_A}{4\,T_R\,N_f} + \frac{C_F}{2} - \frac{2 C^2_F}{C_A}$ and $T_R = \h$. The second equation shows the projector operator for $N_c = N_f = 3$.

We need to find the initial conditions in $\omega$-representation. For $G\Lb\omega, Y_0,\xi\Rb$ we need to re-write in $\omega$-representation \eq{DGLAP3} which takes the form
\beq \label{DGLAP10}
G^D\Lb\omega, Y_0,\xi\Rb\,\,=\,\,\frac{\bas\,\varepsilon}{ 2 \,\pi}\frac{1}{ \omega }
\eeq

  The contribution of $G$ to \eq{DGLAP9} for $n^D$ is equal  to (see \eq{DGLAP10})
        \beq\label{NG1}  
       n^D_{G}\,\,=\,\,e^{-\xi}\int^{\epsilon + i \infty}_{\epsilon  - i \infty}\frac{ d \,\omega}{2\,\pi\,i} \,\frac{\bas\,\varepsilon}{ 2 \,\pi}\frac{1}{  \omega }\,\Bigg\{e^{ \omega\,\delta Y\,\,\,+\,\,\bas\, \lambda_{+}(\omega)\,\xi}\,\,+\,\,\rho\,\omega \,\Lb 
       e^{ \omega\,\delta Y\,\,\,+\,\,\bas\, \lambda_{+}(\omega)\,\xi}\,\,-\,\, e^{ \omega\,\delta Y\,\,\,+\,\,\bas\, \lambda_{-}(\omega)\,\xi}    \Rb\Bigg\}         \eeq
    
    For $\lambda_{+}\Lb \omega\Rb$ we use $\lambda_{\rm approx} \,=\,1/\omega - 1$. This substitution simplifies the first term which takes the following form:
      \beq \label{NG2}
    n^D_1\,=\,\,\frac{\bas }{2 \pi} \Lb x^2_{01}\,\bar{Q}^2\Lb Y_0,b\Rb\Rb\,e^{- \bas \,\xi}\,I_0\Lb 2 \sqrt{\bas \,\delta Y\,\xi}\Rb
    \eeq    
    
    The term with $\lambda_{-}$ decreases at large $\xi$ but it has singularities at $\omega = - k$ with $k=1,2,\dots$. Having this in mind we can find the corrections which could be essential at large $\beta$. In vicinity of $\omega = -1$,   $\lambda_{-}$ takes a form (for $N_c=N_f = 3$)
    \beq \label{NG3}
    \lambda_{-}\Lb \omega\Rb\,\,=\,\,-\, \frac{7}{9}\frac{1}{\omega + 1}
    \eeq
   Replacing $\lambda_{-}$ by \eq{NG3} and $\lambda_{+}$ by $\lambda_{\rm approx}$ we obtain for $n^D$
    \bea \label{NG4}
      n^D\,&=&\,\,\frac{\bas }{2 \pi} \Lb x^2_{01}\,\bar{Q}^2\Lb Y_0,b\Rb\Rb\Bigg\{e^{- \bas \,\xi}\Lb \,I_0\Lb 2 \sqrt{\bas \,\delta Y\,\xi}\Rb \,\,+\,\,\rho \sqrt{\frac{\bas\,\xi}{\delta Y}}\,I_1\Lb 2 \sqrt{\bas\,\delta Y\,\xi}\Rb \Rb\,\,\nn\\
      &+&\,\,\rho e^{- \delta Y} \sqrt{\frac{7 \,\bas\,\xi}{9\,\delta Y}}\,J_1\Lb 2\sqrt{ \frac{7}{9} \bas\,\delta Y\,\xi}\Rb\Bigg\} 
  \eea

    Finally, we need to replace $n^D_{\eq{DLA2}}$ in \eq{DLA3}, by $n^D_{\eq{NG4}}$.   We compare the resulting formula  we compare  with the  42 experimental points (see Ref.\cite{HERADATA}) with $\beta \,\leq\,0.056$, having  the same  other kinematic variables  as in  sections III  and IV. We obtain the description with $\chi^2 = 97$ and $\bas = 0.0549$. From this comparison we can conclude that the improvement  which is related to the full kernels of the DGLAP equation,  is not
    essential for the description of the experimental data.  
   ~

   ~
   
   ~
   
  \section{Conclusions}
   In this paper we developed the DGLAP evolution in the region of small $x_\pom$ and small $\beta$ for the diffractive production in DIS in two different cases: the elastic amplitude at  $Y=Y_0$ is small and outside of the saturation region; and the saturation effects are essential in $N_{\rm el}$.Both approachers can describe the available experimental data but the price for this description is a small value of the QCD coupling ($\bas \sim 0.1$). The conclusion from these attempts  is that the system of produced gluons are dilute,  even at  small values of $\beta$ as $\beta = 0.0056$.
   
   At first sight,  the origin of this dilute system of gluons,  stems from the small value of the elastic amplitude at $\tau_0 = r^2 Q^2_s\Lb Y_0,b\Rb \,=\,1$. In  Refs. \cite{CLP,CLMP} we found that $N_{\rm el}\Lb \tau_0 =1\Rb = 0.1 - 0.3$. However, we showed that from the master equations (see \eq{EQ4}), the initial condition for the gluons density of the emitted gluons has 
   the form: $ \h \bas\,r^2 \bar{Q}^2_s\Lb Y_0,b\Rb$\footnote{This fact was first shown in Ref.\cite{MAR} in framework of the direct calculations in perturbative QCD.}, where $\bar{Q}^2_s\Lb Y_0,b\Rb$ is given by  the integral over all $\tau$ (distances) in  \eq{IC4}. This integral cannot be estimated only from $N_{\rm el}$ in the vicinity of the saturation scale, and  we used the saturation models of Refs. \cite{CLP,CLMP} to its estimate it. The value, which we obtain $\bar{Q}^2_s\Lb Y_0,b\Rb = 1.2 \,Q^2_S\Lb Y_0,b\Rb$, shows that the smallness of $N_{\rm el}$ at $\tau_0 = 1$,  does not induce a small
   initial gluon density. 
   
   Hence, the only reason for a  small  density of emitted gluons we is the small probabilities of their emission, which is characterized by small $\bas$.   The only reasonable explanation why we obtain a small value of the coupling, stems from the importance of the next-to-leading corrections in DGLAP and BFKL evolution, which we are planning to approach in the future publications.

  \section{Acknowledgements}
   We thank our colleagues at Tel Aviv university and UTFSM for
 encouraging discussions. Our special thanks go to Asher Gotsman, 
 Alex Kovner and Misha Lublinsky for elucidating discussions on the
 subject of this paper.

  This research was supported by the BSF grant   2012124, by 
   Proyecto Basal FB 0821(Chile),  Fondecyt (Chile) grants  
 1140842, 1170319 and 1180118 and by   CONICYT grant PIA ACT1406. 
\appendix

\begin{boldmath}
 \section{Solution for $\bas \delta Y\,\xi \sim 1$. } 
 \end{boldmath}
 
 In this appendix we  obtain the  cross section
 for the diffractive production in the kinematic region, where
 $r^{2}Q^{2}_{s}(Y_{0},b)\approx 1$ and  $\dy\ll1$, but do not use
  the assumption, that $\bas \xi \gg 1$ ,which we used
 in section IIC-3. As in this section we are dealing with
 the integral
\begin{equation}\label{repre1}
  \displaystyle{\sum_{n=0}^{\infty}\frac{(\bas \dy)^{n} }{n!} \intee
 \dfrac{1}{\gamma ^{n}}\dfrac{\ea}{\gamma-\gga}.      }
\end{equation}
We wish to calculate this integral using the special function in
 the most compact and economic way. First, we note that  for $n=0$,
 we can take the integral closing the contour of the integration
 over the pole $\gamma = \tilde{\gamma}$.
 For each $n\geq 1$,  we can write the contribution as the convolution integral
 \begin{equation}\label{repreintegral}
\displaystyle{ \intee \dfrac{1}{\gamma ^{n}}\dfrac{\ea}{\gamma-\gga}}\,\,  =\,\,\displaystyle{e^{(\tilde{\gamma}-1)\,\xi} \int_{x}^{1}\dfrac{1}{\Gamma(n)}(-\ln(t))^{n-1} t^{\gga} \dfrac{dt}{t} },
\end{equation}
with $x=e^{-\xi}$.

Plugging   \eqref{repreintegral} into \eqref{repre1} we obtain the following
 expression
\begin{equation}\label{repre2}
n^{D}(Y,Y_{0},\xi,b)\,= \,\,\,e^{(\gga-1)\xi}\left(1+\displaystyle{ \int_{x}^{1}\dfrac{\aal \dy}{\sqrt{\aal \dy \ln(1/t)}}I_{1}(2\sqrt{\aal \dy \ln(1/t)}) t^{\gga}\dfrac{dt}{t}}\right).  
\end{equation}

The next step is to express the integral on the r.h.s. in term of Lommel's
 function. Using that
\begin{equation}\label{entrebessel}
  \dfrac{d}{dt}I_{0}(2\sqrt{\aal \dy \ln(1/t)})=-\dfrac{\aal \dy}{\sqrt{\aal \dy \ln(1/t)}}I_{1}(2\sqrt{\aal \dy \ln(1/t)})\dfrac{1}{t}.
\end{equation} 
 the expression in \eqref{repre2} can be written as follows
\begin{equation}\label{repre3}
n^{D}(Y,Y_{0},\xi,b)\,=\,\,e^{(\gga-1)\xi}\left(1- \displaystyle{ \int_{x}^{1} \dfrac{d}{dt}I_{0}(2\sqrt{\aal \dy \ln(1/t)}) t^{\gga} dt }\right).
\end{equation}
Taking integration by parts into \eqref{repre3} we obtain
\begin{equation}\label{repre4}
n^{D}(Y,Y_{0},\xi,b)\,=\,\,e^{-\xi} I_{0}(2\sqrt{\aal \dy \xi})+\gga e^{(\gga-1)\xi}  \int_{x}^{1} I_{0}(2\sqrt{\aal \dy \ln(1/t)})\, t^{\gga} \dfrac{dt}{t} .
\end{equation}
Using $u=\sqrt{\ln(1/t)/\ln(1/x)}$, the integral in the above expression 
 can be  rewritten as
\begin{equation}\label{repre5}
n^{D}(Y,Y_{0},\xi,b)=e^{-\xi} I_{0}(2\sqrt{\aal \dy \xi})+2\gga \xi e^{-\xi}  \int_{0}^{1} e^{\gga \xi(1- u^{2})}I_{0}(2\sqrt{\aal \dy\,\xi}\, u) u du .
\end{equation}
Introducing  $z=2i\sqrt{\aal \dy \xi}$, $w=-2i\gga \xi$ into \eqref{repre5} yields the following  representation
 \begin{equation}\label{repre6}
n^{D}(Y,Y_{0},\xi,b)=\,\,\Bigg(e^{-\xi} I_{0}(2\sqrt{\aal \dy \xi})\,-\, e^{-\xi}\dfrac{1}{i}\left( U_{1}(w,z)\,\,+\,\,i\,U_{2}(w,z)  \right)\Bigg),
\end{equation}
where $U_{\nu}(w,z)$ denotes the Lommel function of two variables. The 
series representation
\begin{equation}\label{series}
\begin{array}{rcl}
U_{1}(w,z)&=&i\displaystyle{ \sum_{m=0}^{\infty} \left(- \dfrac{\gga \xi}{\sqrt{\aal \dy \xi}} \right)^{2m+1}I_{2m+1}(2\sqrt{\aal \dy \xi})  }\\
U_{2}(w,z)&=&-\displaystyle{ \sum_{m=0}^{\infty} \left(- \dfrac{\gga \xi}{\sqrt{\aal \dy \xi}} \right)^{2m+2}I_{2m+2}(2\sqrt{\aal \dy \xi})  }
\end{array}
\end{equation}
becomes
 \begin{equation}\label{repre7}
n^{D}(Y,Y_{0},\xi,b)=\,\,\Lb e^{-\xi} I_{0}(2\sqrt{\aal \dy \xi})\,+\, e^{-\xi}\displaystyle{\sum_{m=1}^{\infty}\left( \dfrac{\gga \xi}{\sqrt{\aal \dy \xi}} \right)^{m} I_{m}(2\sqrt{\aal \dy \xi})  }\Rb.
\end{equation}
As $\xi\gg 1$, the expression \eqref{repre7} it is not a suitable
 representation for the asymptotic analysis. This problem is 
resolved considering the generating function of the Bessel 
functions that  can be  written as follows
\begin{equation}\label{genera}
  e^{(z/2)\left( t+t^{-1} \right)}=\displaystyle{ \sum_{m=-\infty}^{\infty}t^{m}I_{m}(z)  }
\end{equation}
Introducing \eq{genera} into \eq{repre7}, and using that 
 $I_{m}(z)=I_{-m}(z)$, we obtain
\beq \label{repre}
n^{D}(Y,Y_{0},\xi,b)=\,\,\Lb e^{(\gga-1) \xi +\dfrac{\aal \dy}{\gga}}\,\,-\,\,\displaystyle{e^{-\xi} \sum_{m=1}^{\infty}\left( \dfrac{1}{\gga}\sqrt{\dfrac{\aal \dy}{\xi}}  \right)^{m} I_{m}(2\sqrt{\aal \dy \xi}).}\Rb
\eeq
At large $\aal \dy \xi $,  \eq{repre} takes the form
\begin{equation}\label{asimrepre}
n^{D}(Y,Y_{0},\xi,b)=\,\,\Lb e^{(\gga-1) \xi +\dfrac{\aal \dy}{\gga}}\,-\,\frac{1}{ 2\sqrt{\pi}}\displaystyle{ \left( \dfrac{(\aal \dy)}{\xi^3}  \right)^{1/4}\,e^{-\xi+2\sqrt{\aal \dy\xi}}}\Rb
\end{equation}
If
 we  consider
$\sqrt{\aal \dy/\xi}$ being much larger than $\gga$, one can see that this equation coincides with \eq{V61}

~

~

\begin{boldmath}
 \section{ Initial conditions for the gluon emissions: terms that are proportional to $N^2_{el}$}. 
 \end{boldmath}
 In this appendix we consider \eq{IC00} in more details.   
Using $x^{2}_{ij}$ ,which  denotes $\Vert \vec{x}_{ij} \Vert^{2}$, and $\vec{x}_{20}$ as the  variable in the integration., we write
\begin{equation}\label{dif}
  x_{12}^{2}=(x_{10}-x_{20})\cdot (x_{10}-x_{20})=\xx -2 \Vert x_{10}\Vert \Vert x_{20}\Vert\cos(\theta)+\yy,
\end{equation}
where  $\theta$ is the  angle between  $\vec{x}_{10}$ and $\vec{x}_{20}$.
In the integral of \eq{IC00}  $\Omega_{1}$ is assumed  to be $ N_{el}(Y_{0},x_{ij},b)\sim (\sat^{2}x^{2}_{ij})^{\ga}$ since we are calculating this integral in the kinematic region $x^2_{01}\,Q^2_s\Lb Y_0,b\Rb\,<\,1$. From this form of $N_{el}$ we obtain

\begin{subequations}\label{funciones1}
\begin{align}
\dfrac{\neyy}{\nexx}&= (\yy/\xx)^{\ga},\label{parte1}\\
\dfrac{\nexy}{\nexx}&= \left( 1-2\dfrac{\Vert x_{20} \Vert}{\Vert x_{10}\Vert} \cos(\theta) +\dfrac{\yy}{\xx} \right)^{\ga}.\label{parte2}
\end{align}
\end{subequations}
Using polar coordinate for integrating over $\vec{x}_{02}$ and introducing a new 
radial variable considering $r^{2}=\yy/\xx$,  we obtain \eq{IC00} in the form
\begin{equation}\label{integral1}
I_{1}=\as f_{1}\nexxp+\as f_{2}\nexxs +\as f_{3}\nexxt,
\end{equation}
with
\begin{subequations}\label{descomposicion1}
\begin{align}
f_{1}&=\dfrac{1}{2\pi}\displaystyle{ \int_{0}^{1}\dfrac{dr^{2}}{r^{2}} \int_{0}^{\pi}\dfrac{d\theta}{1-2r\cos(\theta)+r^{2}}\left[ (r^{2})^{2\ga}+(1-2r\cos(\theta) +r^{2})^{2\ga}-1+2(r^{2})^{\ga}(1-2r\cos(\theta) +r^{2})^{\ga} \right]  },\label{potencia2}\\
f_{2}&=\dfrac{1}{2\pi}\displaystyle{ \int_{0}^{1}\dfrac{dr^{2}}{r^{2}} \int_{0}^{\pi}\dfrac{d\theta}{1-2r\cos(\theta)+r^{2}}\left[(r^{2})^{2\ga}(1-2r\cos(\theta) +r^{2})^{\ga}+(r^{2})^{\ga}(1-2r\cos(\theta)+r^{2}  )^{2\ga}    \right]},\label{potencia3}\\
f_{3}&=\dfrac{1}{2\pi}\displaystyle{ \int_{0}^{1}\dfrac{dr^{2}}{r^{2}} \int_{0}^{\pi}\dfrac{d\theta}{1-2r\cos(\theta)+r^{2}}\left[ (r^{2})^{2\ga}(1-2r\cos(\theta) + r^{2})^{2\ga} \right]},\label{potencia4}
\end{align}
\end{subequations}
It should be noted that all integrals in the above equations are convergent ones.
 Their values can be  obtained through the hypergeometric functions. Specifically, in the procedure of the calculation we use the following relations:
\begin{subequations}\label{relaciones}
\begin{align}
\displaystyle{\int_{0}^{\pi}\dfrac{\sin^{(2\mu -1)}(\theta)}{(1-2a\cos(\theta) +a^{2})^{\nu}  }}&=B(\nu,1/2)\ _{2}F_{1}(\nu,\nu-\mu+(1/2);a^{2}),\qquad (a^{2}<1)\label{rel1}\\
\displaystyle{ \int\dfrac{x^{b}}{1-x} dx }&=\dfrac{x^{b+1}}{b+1}\ _{2}F_{1}(1,b+1,b+2;x),\label{rel2}\\
\displaystyle{ \int x^{b} _{2}F_{1}(a,a,1,x)  }&=\dfrac{x^{b+1}}{b+1}\ _{3}F_{2}(\{a,a,b+1\},\{1,b+2\};x),\label{rel3}\\
\displaystyle{ _{2}F_{1}(1,1,1,x)  }&=\dfrac{1}{1-x},\label{rel4}\\
 \displaystyle{ \lim_{x\to 1^{-}} \dfrac{_{2}F_{1}(a,b,c,x)}{-\ln(1-x)} }&=\displaystyle{ \dfrac{\Gamma(a+b)}{\Gamma(a)\Gamma(b)}  }\qquad (c=a+b)\label{rel5}.
\end{align}
\end{subequations}
Here $B$ denotes the Beta function whereas $_{p}F_{q}$ correspond to the (generalized) hypergeometric function represented by the following series:
\begin{equation}\label{hyper}
_{p}F_{q}(\{p_{1},\dots,p_{m}\},\{q_{1},\dots,q_{n}\};x)=\displaystyle{\sum_{k=0}^{\infty}\dfrac{\prod_{j=1}^{m}(p_{j})_{k}}{\prod_{j=1}^{n}(q_{j})_{k}   }\dfrac{x^{k}}{k!}   }\qquad (\vert x\vert<1),
\end{equation}
with $(a)_{k}=\Gamma(a+k)/\Gamma(a)$ (see Ref.\cite{RY} for more details for integral representations of  \eqref{relaciones} and for the special functions, respectively).  It should be stressed that the equation of \eqref{rel5} shows that we do not face any logarithmic  infrared divergency  in our procedure of the estimates.
 Considering $\mu=1/2$  in  \eqref{rel1} (and knowing $B(1/2,1/2)=\pi$),  the closed expression for angular integrals in \eqref{potencia2} is  obtained. Finally, $f_1$ takes the form
\begin{equation}\label{f1}
f_{1}=\dfrac{1}{2}\displaystyle{\int_{0}^{1}dz\left[ \dfrac{z^{2\ga-1}-1}{1-z}+\dfrac{1}{z}\ ( _{2}F_{1}(1-2\ga,1-2\ga,1;z)-1)+2 z^{\ga-1}\ _{2}F_{1}(1-\ga,1-\ga,1;z)  \right],\ \textrm{(with $z=r^{2}$)}   }
\end{equation}
where
\begin{equation}\label{integralesseparadas}
\begin{array}{rcl}
\displaystyle{\int_{0}^{1}dz\dfrac{z^{2\ga-1}-1}{1-z}}&=& \psi(1)-\psi(2\ga),\\
\displaystyle{\int_{0}^{1}dz \dfrac{1}{z}\ ( _{2}F_{1}(1-2\ga,1-2\ga,1;z)-1)}&=&\displaystyle{ \sum_{k=1}^{\infty} \dfrac{((1-2\ga)_{k})^{2}}{(k!)^{2}k} },\\
\displaystyle{\int_{0}^{1}dz \  _{2}F_{1}(1-\ga,1-\ga,1;z)z^{\ga-1}}&=&\displaystyle{ \sum_{k=0}^{\infty} \dfrac{((1-\ga)_{k})^{2}}{(k!)^{2}(\ga+k)}},
\end{array}
\end{equation}
and therefore
\begin{equation}\label{primero}
\begin{array}{rcl}
f_{1}&=&\displaystyle{\dfrac{ \psi(1)-\psi(2\ga)}{2}+ \dfrac{1}{2}\sum_{k=1}^{\infty} \dfrac{((1-2\ga)_{k})^{2}}{(k!)^{2}k}+ \sum_{k=0}^{\infty} \dfrac{((1-\ga)_{k})^{2}}{(k!)^{2}(\ga+k)}    }, \\
     &=&1.59
\end{array}
\end{equation}
Similarly,  using \eqref{relaciones} the terms $f_{2}$ and $f_{3}$  can be reduced to the following forms:
\begin{equation}\label{seguntercer}
\begin{array}{rcl}
f_{2}&=&-\displaystyle{\int_{0}^{1}dz\left[ z^{2\ga-1}\ _{2}F_{1}(1-\ga,1-\ga,1;z)+z^{\ga-1}\ _{2}F_{1}(1-2\ga,1-2\ga,1;z)  \right]  }\\
     &=&-\displaystyle{ \sum_{k=0}^{\infty}\dfrac{((1-\ga)_{k})^{2}}{(k!)^{2}(2\ga+k)}  - \sum_{k=0}^{\infty}\dfrac{((1-2\ga)_{k})^{2}}{(k!)^{2}(\ga+k)}},\\
     &=&-\,2.53.\\
f_{3}&=&\displaystyle{\dfrac{1}{2}\int_{0}^{1}dz \ _{2}F_{1}(1-2\ga,1-2\ga,1;z) z^{2\ga -1}   }\\
     &=&\displaystyle{ \dfrac{1}{2} \sum_{k=0}^{\infty}\dfrac{((1-2\ga)_{k})^{2}}{(k!)^{2}(2\ga+k)}}\\
     &=&0.413876
\end{array}
\end{equation}
Hence,  the integral $I_1$ has the value
\begin{equation}\label{rama1pre}
I_{1}=1.59\,\bas\,\nexxp\,- \,2.53\bas\, \nexxs \,+\,0.414\,\bas\,\nexxt,
\end{equation}
and since $\tau_{0}\lessapprox 1$, we can safely  assume that the value of $I_1$  is given by
\begin{equation}\label{rama1}
I_{1}=1.59\,\bas\,\nexxp
\end{equation}

We now  discuss the contribution of the following integral 
\bea \label{B2}
 I_2 \,\,&=&\,\,\,\frac{\bas}{ 2 \,\pi}\,
 x^2_{10}\, \int_{x^2_{01}}\, \frac{d^2 \vec{x}_{02}}{\vec{x}^2_{02}\,\vec{x}^2_{12}}\Bigg[ \Bigg\{ -\,  N^2_{\rm el} \Lb Y_0, \vec{x}_{01},\vec{b} \Rb \nn\\
 & +&\,\,
\Bigg(\, N_{\rm el} \Lb Y_0, \vec{x}_{02},\vec{b}\Rb \,+\,N_{\rm el} \Lb Y_0, \vec{x}_{12},\vec{b} \Rb
-\, N_{\rm el} \Lb Y_0, \vec{x}_{02},\vec{b} \Rb\, N_{\rm el} \Lb Y_0, \vec{x}_{12},\vec{b} - \h \vec{x}_{02}\Rb\Bigg)^2 \Bigg\}\nn\\
&-&\,\,\Bigg\{ 2\,  N_{\rm el} \Lb Y_0, \vec{x}_{02},\vec{b} \Rb \,-\, N^2_{\rm el} \Lb Y_0, \vec{x}_{02},\vec{b} \Rb\Bigg\}^2\Bigg] \eea

It is clear that for the integral $I_2$ we have the same expansion as for $I_1$: viz.
\begin{equation}\label{integral21}
I_{2}\,\,=\,\,\bas\, \tilde{f}_{1}\nexxp\,+\,\bas \,\tilde{f}_{2}\nexxs \,+\,\bas\, \tilde{f}_{3}\nexxt,
\end{equation}
where each $\tilde{f}_{i}$ is described similarly as \eqref{descomposicion1} but for  $r$ being in the range  $1\,\leq\,r\,\leq\,\,\infty$ .  Changing  the variable $r \to z$ with $z\,=\,1/r^2$ we obtain
\begin{equation}\label{parte1dom2}
\tilde{f}_{1}=\displaystyle{ \dfrac{1}{2}\int_{0}^{1}dz\left[ \dfrac{z^{1-2\ga}-1}{1-z} +z^{-2\ga}\Lb {}_{2}F_{1}(1-2\ga,1-2\ga,1,z) \,-\,1\Rb+ 2 z^{-2\ga}\Lb{} _{2}F_{1}(1-\ga,1-\ga,1,z) \,-\,1\Rb \right]},
\end{equation}
Taking integrals we obtain
\bea
\hspace{-0.8cm}\tilde{f}_{1}&=&-\,\h\,H_{1-2 \ga }   + \frac{{}_3F_2(1-2 \ga ,1-2 \ga ,1-2 \ga ;1,2-2 \ga ;1)-1}{2(1-2 \ga)} \,+\,
\frac{{} _3F_2(1-2 \ga ,1-\ga ,1-\ga ;1,2-2 \ga ;1)-1}{1-2 \ga }
\eea

where  $ H_{1 - 2 \ga}\,\,=\,\,\psi(1) - \psi( 2(1 - \ga))$ is harmonic number.

For  the numerical value for $\ga$ we obtain that
\beq \label{B3}
\tilde{f}_{1}\,\,=\,\,0.448\,\bas \,\,\,\nexxp
\eeq

Calculating  $\tilde{f}_{1}$ we assumed that the integral of \eq{parte1dom2} is diverged in the region of $\tau_0 \sim 1$ where we can  replace  $ N_{el}(Y_{0},x_{ij},b)$ by $ (\sat^{2}x^{2}_{ij})^{\ga}$ . However, 
for estimates of the term    $\,\bas\, \tilde{f}_{3}\nexxt$ in \eq{integral21} we cannot use this replacement and have to calculate it using the saturation model of Ref.\cite{CLP}.

 The last contribution, which proportional to $N_{\rm el}^2$, stems from \eq{IC3} and \eq{IC4}.In both these equations 
 we integrated over $x^2_{02}$ from $x^2_{02} = 0$ while actually this integration should be for $x^2_{02} \geq x^2_{01}$. Hence we need to subtract the following integral
 \bea \label{B4}
 I_4\,&=&\,\frac{\bas}{2} x^2_{01}\,\int^{x^2_{01}}_0 \, \frac{d x^2_{02}}{\vec{x}^4_{20}}
\Bigg(\,2\, N_{\rm el} \Lb Y_0, \vec{x}_{02},\vec{b} - \h \vec{x}_{02}\Rb \,-\, N^2_{\rm el} \Lb Y_0, \vec{x}_{02},\vec{b} - \h \vec{x}_{02}\Rb\Bigg)^2 \nn\\
&\,\,=\,\,&\bas\Lb \frac{2}{ 2 \ga -1} \nexxp \,-\,\frac{2}{3 \ga -1} \nexxs \,+\,\frac{1}{2 (4 \ga -1)}\nexxt\Rb
\eea

Collecting all contributions we can see that we need to add
\beq \label{B5}
f^{\Sigma}_1\,\,=\,\,-\, 5.62\,\bas\,\nexxp
\eeq

\begin{boldmath}
 \section{Anomalous dimensions of the DGLAP equations. } 
 \end{boldmath}
For the completeness of presentation we include the well know formulae for the anomalous dimensions in the leading order of perturbative QCD (see, for examples, Refs. \cite{KOLEB,EKL}).
\begin{subequations}\label{ogammas}
  \begin{align}
    \gamma_{ff} (\omega) \, = & \,\frac{ C_F }{2 \,N_c}\, \left\{ \frac{3}{2} +
      \frac{1}{(1+\omega) \, (2 + \omega)} - 2 \, \psi (\omega + 2) +
      2 \, \psi (1) \right\} \\
    \gamma_{gf} (\omega) \, = & \,\,\frac{ C_F }{2 \,N_c}\, \left\{ \frac{1}{2+\omega} +
      \frac{2}{\omega \, (1+\omega)} \right\} \\
    \gamma_{fg} (\omega) \, = & \, \frac{ N_f }{2 \,N_c}\, \left\{ \frac{1}{1+\omega} -
      \frac{2}{(2+ \omega) \, (3+\omega)} \right\} \\
    \gamma_{gg} (\omega) \, = & \, \frac{11 \, N_c - 2 \, N_f}{12\,N_c}
   +  \left\{ \frac{1}{\omega \, (1+\omega)} +
      \frac{1}{(2+ \omega) \, (3+\omega)} - \psi (\omega + 2) + \psi
      (1) \right\}
  \end{align}
\end{subequations}
where $\psi (w) = \Gamma'(w)/\Gamma (w)$ is the digamma function. Note
that $\psi (1) = - \gamma_E$ where $\gamma_E$ is Euler's constant.  $N_f$ is the number of the fermions (quarks) and $C_F = (N^2_c - 1)/(2\,N_c)$.

~

~


 \section{Energy evolution of the saturation scale. } 

In this appendix we discuss \eq{DLA5} as well as the fact that ${\rm Const}$ in \eq{DLA6} is proportional to $\ln\Lb x^2_{01} Q^2_s\Lb Y\Rb\Rb$. \eq{DLA5} can be written as
\beq \label{D1}
\ln\Lb Q^2_s\Lb \delta Y\Rb\Big{/}Q^2_s\Lb Y_0\Rb\Rb\,\,=\,\,\bas \,\kappa \,\delta Y\,-\,\frac{3}{2\,\bar{\gamma}}\,\ln\Lb \delta Y\Rb
\eeq
First term was derived in Ref.\cite{GLR} and the second in Ref.\cite{MUT}.
Let us introduce a new variable $z_0 = \ln\Lb x^2_{01}\,Q^2_s\Lb Y_0\Rb\Rb \,+\,\bas \,\kappa\,\delta Y$. \eq{D1} stems from two observations. First, that for $z_0 \,<\,0$ we can use for $n^D$ the solution of the linear equation:
\bea \label{D2}
&&n^D\Lb z_0 \,<\,0\Rb\,=\,\int \frac{d \gamma}{2 \,\pi\,i} n^D_{\rm in}\Lb \gamma; \{\dots\}\Rb\exp\Bigg\{\Lb \chi\Lb \bar{\gamma}\Rb \,+\,\chi'\Lb \bar{\gamma}\Rb\,\Lb \gamma - \ga\Rb  \,+\,\h \chi''\Lb \bar{\gamma}\Rb\,\Lb\gamma - \bar{\gamma}\Rb^2\Rb \,\bas\, \delta Y\,-\,\Lb 1 - \gamma\Rb \xi\Bigg\}\nn\\
&&=\,\,\int \frac{d \gamma}{2 \,\pi\,i} n^D_{\rm in}\Lb \gamma; \{\dots\}\Rb\exp\Bigg\{\bar{\gamma}\,z_0 \, \,+\,\h \chi''\Lb \bar{\gamma}\Rb\,\Lb\gamma - \bar{\gamma}\Rb^2\,\,\bas\, \delta Y\,-\,\Lb \bar{\gamma} - \gamma\Rb\,z_0\Bigg\}\eea

The second observation is that $n^D$ for $z_0\,=\,0$ is not a constant by $n^D\Lb z_0 = 0\Rb \,\xrightarrow{\delta Y \gg 1} 1/\Lb \bas \delta Y\Rb$. This follows directly from semi-classical equations (see, for example,  Ref.\cite{KOLEB}, 4.5.3,Eq.4.187). Since on the critical line $z_0=0$ we can use the solution of \eq{D2} we see that $n^D\Lb \gamma \{\dots\}\Rb \,\,=\,\,\Lb 2/\bas\Rb \,\Lb \bar{\gamma} - \gamma\Rb$. For $\xi \,\neq\,\bas\,\kappa\,\delta Y$, the integral of \eq{D2} taken by the method of steepest descent gives:
\beq \label{D6}
n^D\Lb z_0 \,<\,0\Rb\,=\,2\sqrt{\frac{2\,\pi}{\chi''\Lb \bar{\gamma}\Rb}}\,\frac{z_0}{\Lb\bas \,\delta Y\Rb^{3/2}}\,e^{\bar{\gamma} \,z_0}
\eeq
In \eq{D6} we consider $\delta Y \,\gg\,z_0$ and neglect the terms which rate proportional to $z^2_0/\delta Y$.

We need to taker \eq{D1} for the saturation momentum to compensate this behaviour. Plugging the new saturation momentum in \eq{D2} we obtain that in the vicinity of the new saturation momentum, 
$n^D$ takes the following form
\beq \label{D7}
n^D\,\,=\,\,{\rm Const}\,\ln\Lb x^2_{01}\,Q^2\Lb Y\Rb\Rb \,\Lb x^2_{01}\,Q^2_s\Lb \delta Y\Rb\Rb^{\bar{\gamma}}
\eeq
\eq{D7} reproduces  $n^D$ in the form obtained in Ref.\cite{MUMU}.

\end{document}